\documentclass[12pt, reqno]{amsart}
\usepackage{lscape,amssymb, amsmath,verbatim,graphicx,xcolor,url,booktabs,longtable,paralist}
\usepackage{epsfig,subfigure,float,setspace,parskip}
\usepackage[pdftex,colorlinks = true]{hyperref}
\hypersetup{citecolor=darkblue,linkcolor=darkblue,urlcolor=darkblue}
\definecolor{darkblue}{rgb}{0,0,.6}
\usepackage[nolists]{endfloat}

\usepackage{multicol}
\usepackage{epsfig,microtype}
\usepackage[margin=1in]{geometry}
\newtheorem{assumption}{Assumption}[section]

\newcommand{\beps}{{\mbox{\boldmath $\epsilon$}}}

\newcommand{\intt}{\int\hspace{-.2cm}\int}

\newtheorem{theorem}{Theorem}[section]
\newtheorem{lemma}{Lemma}[section]

\allowdisplaybreaks
\def\beq{\begin{equation}}
\def\eeq{\end{equation}}

\graphicspath{{plots/}}

\begin{document}

\title[Long run covariance bandwidth selection]{A plug-in bandwidth selection procedure for long run covariance estimation with stationary functional time series}

\author{Gregory Rice}
\address{ Department of Statistics and Actuarial Science, University of Waterloo, Waterloo, ON, Canada}

\author{Han Lin Shang}
\address{ Research School of Finance, Actuarial Studies and Statistics, Australian National University, Canberra, ACT 2601, Australia}


\maketitle

\begin{abstract}

In arenas of application including environmental science, economics, and medicine, it is increasingly common to consider time series of curves or functions. Many inferential procedures employed in the analysis of such data involve the long run covariance function or operator, which is analogous to the long run covariance matrix familiar to finite dimensional time series analysis and econometrics. This function may be naturally estimated using a smoothed periodogram type estimator evaluated at frequency zero that relies crucially on the choice of a bandwidth parameter. Motivated by a number of prior contributions in the finite dimensional setting, we propose a bandwidth selection method that aims to minimize the estimator's asymptotic mean squared normed error (AMSNE) in $L^2[0,1]^2$. As the AMSNE depends on unknown population quantities including the long run covariance function itself, estimates for these are plugged in in an initial step after which the estimated AMSNE can be minimized to produce an empirical optimal bandwidth. We show that the bandwidth produced in this way is asymptotically consistent with the AMSNE optimal bandwidth, with quantifiable rates, under mild stationarity and moment conditions. These results and the efficacy of the proposed methodology are evaluated by means of a comprehensive simulation study, from which we can offer practical advice on how to select the bandwidth parameter in this setting.

\vspace{.1in}


\end{abstract}

\section{Introduction}

Functional time series analysis has grown substantially in recent years in order to provide methodology for studying functional data objects that are obtained sequentially over time. Perhaps the most typical way in which such data arises is when long, dense records of a continuous time phenomena are segmented into collections of curves, e.g. high frequency records of pollution levels that are segmented to form daily pollution curves, or records of tick-by-tick asset price data that may be used to construct intraday price or return curves. Other examples include sequentially observed summary functions that describe physical phenomena, as in functional magnetic resonance imaging, in which functions describing blood flow in the brain are computed over time. We refer the reader to \cite{ferraty:vieu:2006} and \cite{hormann:kokoszka:2010} for an overview of the fields of functional data analysis and functional time series analysis, and to \cite{horvath:kokoszka:rice:2014,panaretos:tavakoli:2015,aue:norinho:hormann:2014,aston:kirch:2012,aston:kirch:2012AAS,horvath:kokoszka:reeder:2012,torgovitski:2016,hormann:kidzinski:hallin:2015,zhang:2016} for a sample of recent contributions that focus on inference, and dimension reduction with functional time series data.

Many of the inference and dimension reduction procedures introduced in the above cited papers are based on the second order properties of the sample mean function of functional time series, and hence naturally involve the estimation of the long run covariance function, or corresponding long run covariance operator, of the functional time series. Long run covariance and spectral density estimation enjoys a vast literature in the case of finite dimensional time series, beginning with the seminal work of Bartlett \cite{bartlett:1946} and Parzen \cite{parzen:1957}, and still the most commonly used techniques resort to smoothing the periodogram by employing a smoothing weight function and a bandwidth parameter. Data driven bandwidth selection methods in this setting have received a great deal of attention; see \cite{andrews:1991,buhlmann:1996,newey:west:1987,newey:west:1994,politis:2011,hirukawa:2010}. Roughly speaking, the methods proposed in each of these papers aim to select the bandwidth that minimizes the asymptotic mean squared error (AMSE) of the estimator. Since the AMSE involves the unknown value of the long run variance and higher order derivatives of the spectral density, it is proposed that estimates of these quantities be ``plugged in" to the expression for the AMSE, after which an approximately optimal bandwidth in terms of minimizing this estimated AMSE can be calculated.

Horv\'ath et al \cite{horvath:kokoszka:reeder:2012}, and Panaretos and Tavakoli \cite{panaretos:tavakoli:2012} define analogous smoothed periodogram type estimates of the long run covariance and spectral density operators for functional time series, however the problem of bandwidth selection has been only lightly investigated in this setting. Horv\'ath et al \cite{horvath:rice:whipple:2014} propose an adaptive bandwidth selection algorithm that is designed for the infinite order  ``flat-top" weight function of Politis and Romano \cite{politis:romano:1996} as in \cite{politis:2003adap}, however bandwidth selection methodology for finite order weight functions have not yet been considered. There are a number of benefits to using a finite order weight function in practice. We refer the reader to \cite{marron:wand:1992} for a discussion of these reasons in the context of nonparametric regression that also hold true here, but they include that finite order weight functions (of low enough order) preserve positivity of the resulting estimate of the long run covariance operator, and that, although higher order weight functions are asymptotically more efficient, extremely large sample sizes are needed before they reliably beat their lower order counterparts in practice. Moreover, due to the infinite dimensional nature of functional data, it is unclear how the asymptotic properties of long run covariance estimators are affected when they are adjusted to be positive definite by replacing negative eigenvalues by zero in the diagonalization of the estimated operator, which is a common practice in the finite dimensional case, and this lends further motivation to use a positive definite estimate from the outset.

In this paper, we propose a bandwidth selection method for estimates of the long run covariance function based on finite order weight functions that aims to minimize the estimator's asymptotic mean squared normed error (AMSNE) in $L^2[0,1]^2$. The method involves a similar plug-in step in which unknown quantities appearing in the expression of the AMSNE are estimated using pilot estimates, and the resulting expression is then minimized to produce an optimal bandwidth estimate. We show under simple stationarity, moment, and decay conditions on the autocovariance operators of the functional time series that the proposed method produces a bandwidth that is asymptotically consistent with the optimal bandwidth that minimizes AMSNE with a quantifiable convergence rate. These results and the method itself are thoroughly studied by means of Monte Carlo simulation applied to a number of data generating processes, from which we provide practical advice on how to choose this parameter in applied settings.

The rest of the paper is organized as follows. In Section \ref{main}, we formally define our bandwidth selection method and present its consistency properties. The method is studied and compared to other standard bandwidth choices in Section \ref{simul} by means of a thorough Monte Carlo simulation study. The proofs of all of the results presented in Section \ref{main} are collected in Section \ref{proofs}.

\section{Statement of method and main results}\label{main}

In order to provide a formal definition of the long run covariance function, suppose that $\{X_t(u),\; u\in[0,1]\}_{i \in {\mathbb Z}}$, is a stationary and ergodic functional time series. For example, $X_t(u)$ could be used to denote the density of pollutants in a given city on day $t$ at intraday time $u$, perhaps suitably transformed so that the stationarity assumption is thought to hold. The long-run covariance function is defined as
\begin{equation*}
C(u,s)=\sum_{\ell=-\infty}^\infty\gamma_\ell(u,s) ,\quad \mbox{where  }\;\;\gamma_\ell(u,s)=\mbox{cov}(X_0(u), X_\ell(s)),
\end{equation*}
and is a well defined element of $L^2[0,1]^2$ under mild weak dependence and moment conditions. Via right integration, $C$ defines a Hilbert-Schmidt integral operator on $L^2[0,1]$ given by
\begin{equation*}
c(f)(u) = \int_0^1 C(u,s)f(s)ds,
\end{equation*}
whose eigenvalues and eigenfunctions are related to the dynamic functional principal components defined in \cite{hormann:kidzinski:hallin:2015}, and provide asymptotically optimal finite dimensional representations of the sample mean of dependent functional data.

It is of interest in many applied settings to estimate $C$ from a finite sample $X_1,...,X_T$. Given its definition as a bi-infinite sum, a natural estimator of $C$ is
\begin{align}\label{est-1}
\hat{C}_{h,q}(u,s)=\sum_{\ell=-\infty}^{\infty}W_{q} \left( \frac{\ell}{h} \right) \hat{\gamma}_\ell(u,s),
\end{align}
where $h$ is called the bandwidth parameter,
\begin{align*}
   \hat{\gamma}_\ell(u,s)=\left\{
     \begin{array}{lr}
      \displaystyle \frac{1}{T}\sum_{j=1}^{T-\ell}\left( X_j(u)-\bar{X}(u)\right)\left( X_{j+\ell}(s)-\bar{X}(s)\right),\quad &\ell \ge 0
      \vspace{.3cm} \\
     \displaystyle \frac{1}{T}\sum_{j=1-\ell}^{T}\left( X_j(u)-\bar{X}(u)\right)\left( X_{j+\ell}(s)-\bar{X}(s)\right),\quad &\ell < 0,
     \end{array}
   \right.
\end{align*}
is an estimator of $\gamma_\ell(u,s)$, and $W_q$ is a symmetric weight function with bounded support of order $q$, which is to say that
\begin{align}\label{k-1}
&W_{q}(0)=1,\;W_{q}(u)=W_{q}(-u),\;W_{q}(u)=0\;\;\mbox{if}\;\; |u|>m \;\;\mbox{for some } m>0,\mbox{ and }W_{q} \mbox{ is }\\
&\mbox{continuous on}\;\;[-m,m], \notag
\end{align}
and there exists $w$ satisfying
\begin{align}\label{order}
0< w = \lim_{x\to 0} x^{-q}(W_{q}(x)-1) < \infty.
\end{align}

The estimator in \eqref{est-1} was introduced in \cite{horvath:kokoszka:reeder:2012} and \cite{panaretos:tavakoli:2012}. Only mild conditions must be assumed on the bandwidth parameter $h$ in order for $\hat{C}_{h,q}$ to be a consistent estimator of $C$ in norm, namely that $h=h(T) \to \infty$ as $T\to \infty$ and $h(T)=o(T)$, however its choice can greatly affect the performance of the estimator in finite samples, and hence it is desirable to develop a data driven approach based on the sample at hand to select $h$ in order to minimize the estimation error.

We take as the goal for selecting $h$ to minimize the mean squared error measured by the norm $E\|\hat{C}_{h,q}-C\|^2,$ where $\| \cdot \|$ denotes the standard norm in $L^2[0,1]^2$. It is established in \cite{berkes:horvath:rice:2016} that under mild conditions, which are implied by Assumptions \ref{stat-ergo} and \ref{as-r} below, that
\begin{align}\label{m-fro}
E\|\hat{C}_{h,q}-C\|^2= \frac{h}{T}\Biggl(\|  C\|^2 + \Biggl(\int_0^1 C(u,u)du \Biggl)^2\Biggl)\int_{-\infty}^{\infty}W_q^2(x)dx&+h^{-2{\frak q}}\|{w}C^{(q)}\|^2\\
&+o\Biggl(\frac{h}{T} + h^{-2q}\Biggl), \notag
\end{align}
where
\begin{equation*}
C^{(q)}(u,s)=\sum_{\ell=-\infty}^{\infty}|\ell|^{ q}\gamma_\ell(u,s),
\end{equation*}
and the constant ${w}$ is defined in equation \eqref{order}. In particular, the first two terms on the right hand side of \eqref{m-fro} represent the asymptotically leading terms in the mean squared norm of the error of $\hat{C}_{h,q}$, which we refer to as the AMSNE. This result suggests choosing $h$ to minimize the sum of these two terms, which, by a simple calculus, amounts to choosing $h$ to be
\begin{align}\label{h-opt}
h_{{\it opt}}= {\frak c}_0T^{1/(1+2{\frak q})},
\end{align}
where
\begin{equation*}
{\frak c}_0=\left({ q}\|{C^{(q)}}\|^2\right)^{1/(1+2{ q})}\left(\left(\| C\|^2 + \left(\int_0^1 C(u,u)du \right)^2\right)\int_{-\infty}^\infty W_q^2(x)dx   \right)^{-1/(1+2{ q})}.
\end{equation*}
Of course, the obvious crux here as pointed out in \cite{parzen:1957,buhlmann:1996} is that the quantities involving $C^{(q)}$ and $C$ in \eqref{h-opt} are unknown and include what we are trying to estimate in the first place. However, this motivates the following method to select $h$:

{\it ``Plug-in" Bandwidth selection method: }
\begin{enumerate}

\item Compute pilot estimates of $C^{(p)}$, for $p=0,q$:
\begin{align}\label{cp-def}
\hat{C}^{(p)}_{h_1,q_1}(u,s)=\sum_{\ell=-\infty}^{\infty}W_{q_1}\left( \frac{\ell}{h_1} \right) |\ell|^{p} \hat{\gamma}_\ell(u,s),
\end{align}

that utilize an initial bandwidth choice $h_1=h_1(T)$, and weight function $W_{q_1}$ of order $q_1$.

\item Estimate ${\frak c}_0$ by
$$
\hat{\frak c}_0(h_1,q_1)=\left({ q}\|{\hat{C}^{(q)}_{h_1,q_1}}\|^2\right)^{1/(1+2{ q})}\left(\left(\| \hat{C}_{h_1,q_1}\|^2 + \left(\int_0^1 \hat{C}_{h_1,q_1}(u,u)du \right)^2\right)\int_{-\infty}^\infty W_q^2(x)dx   \right)^{-1/(1+2{ q})}.
$$

\item Use the bandwidth
\begin{align}\label{hopt-def}
\hat{h}_{opt}(h_1,q_1)= \hat{\frak c}_0 (h_1,q_1) T^{1/(1+2{ q})}
\end{align}
 in the definition of $\hat{C}_{h,q}$ in \eqref{est-1}.
\end{enumerate}

The above method defines a direct functional analog of the method of Newey and West \cite{newey:west:1994} that intends to adapt the bandwidth to reflect the underlying dependence of the functional time series. When the data are only weakly correlated, pilot estimates for $\|\hat{C}^{(q)}\|$ are expected to be close to zero, resulting in a small bandwidth that reduces the variance of $\hat{C}$. Under strong dependence, the empirical bandwidth should be large, and this serves to reduce the estimation bias that dominates in this case. The multiplication by the power of $T$ ensures that the bandwidth has the optimal asymptotic rate corresponding to the order of the weight function used in the estimator. Under the following assumptions, we may derive convergence rates of $\hat{h}_{opt}$ to $h_{opt}$ that imply consistency of the empirical optimal bandwidth, and also inform the choice of the values for the order and bandwidth of the pilot estimates.

\begin{assumption}\label{stat-ergo}
The sequence $\{X_t(u),\; u \in[0,1]\}_{i \in {\mathbb Z}}$ is an $L^p-$m approximable Bernoulli shift in $L^2[0,1]$, satisfying

\begin{itemize}
\item  $X_t=g(\epsilon_t,\epsilon_{t-1},...)$ for some measurable function $g:S^\infty \mapsto L^2[0,1]$ and iid random variables $\epsilon_t$ $-\infty<t<\infty,$ with values in a  measurable space $S$.

\item  $E||X_0||^{4+\delta} < \infty $ for some $\delta >0,$ and

\item $\{X_t(u),\; u \in[0,1]\}_{i \in {\mathbb Z}}$ can be approximated by $m\mbox{-dependent}\;\mbox{sequences}$ $X_{t,m}=g(\epsilon_t,\epsilon_{t-1},...,\epsilon_{t-m+1},\beps_{t,m}^*)$, with $\beps^*_{t,m}=(\epsilon_{t,m, t-m}^*,\epsilon_{t,m, t-m-1}^*,\ldots)$,
where the $\epsilon^*_{t,m,k}\mbox{'s}$ are independent copies of $\epsilon_0$, independent of $\{\epsilon_i, -\infty<i<\infty\}$, such that \\
$(E\|X_0-X_{0,m}\|^4)^{1/4}=O(m^{-\rho}),$ for some $\rho>4$.
\end{itemize}

\end{assumption}

\begin{assumption}[$r$]\label{as-r}
There exists $r \ge q$ such that
\begin{align}\label{as-r-c}
\sum_{\ell=-\infty}^\infty (1+|\ell|)^{q+r} \|\gamma_\ell\| < \infty.
\end{align}
\end{assumption}

Assumptions \ref{stat-ergo} and \ref{as-r} compare to the assumptions in \cite{buhlmann:1996,politis:2003adap,politis:2011,politis:romano:1996} in the scalar case, and roughly define that the functional time series $X_t(u)$ is weakly dependent with quantifiable rates of decay of the autocovariance functions. The conditions in Assumption \ref{stat-ergo} and \eqref{as-r-c} can be easily verified for the stationary functional time series models available in the literature to date, which are all based on simple structural equations. These include the functional ARMA model, see \cite{bosq:2000}, and nonlinear models such as the functional ARCH and GARCH models, see \cite{hormann:horvath:reeder:2012} and \cite{aue:horvath:pellatt:2016}, along with transformations of such processes.

Assumption \ref{stat-ergo} is practically equivalent to the main Assumption in \cite{berkes:horvath:rice:2016}, and is needed in this context in order to bound the size of sums of integrals of functional versions of cumulants. If $a_\ell(t,s)=EX_0(t)X_\ell(s)$, then one can define the fourth order cumulant function as
\begin{align*}
\Psi_{\ell,r,p}(t,s)=E[X_0(t)&X_\ell(s)X_r(t)X_p(s)] \\
 &- a_\ell(t,s)a_{p-r}(t,s)- a_r(t,t)a_{p-\ell}(s,s)- a_p(t,s)a_{r-\ell}(t,s). \notag
\end{align*}
Assumption \ref{stat-ergo} above could be replaced with the condition that
$$
\sum_{i,j,k=-\infty}^{\infty} \Biggl|\intt {  { \Psi_{i,j,k}(t,s)}} dtds\Biggl| < \infty,
$$
which is similar to traditional cumulant summability conditions for scalar time series, see \cite{brillinger:1975}, and holds under various mixing conditions. Similar conditions to these are studied in \cite{panaretos:tavakoli:2012} and \cite{zhang:2016}.

The following theorem quantifies the convergence rate of the empirical bandwidth in terms of $r$ in Assumption \ref{as-r} and the bandwidth and order of the pilot estimates.

\begin{theorem}\label{th-1}
Under Assumptions \ref{stat-ergo} and \ref{as-r}, and if $h_1=A T^{\kappa}$, for some constant $A > 0$ and $\kappa<1/(2q+1)$, then
$$
\hat{h}_{opt}(h_1,q_1)= h_{opt}\left(1+O_P(T^{-\beta})\right),
$$
where $\beta = \min\{ 1/2-\kappa(q+1/2), \kappa\alpha \}$, and $\alpha= \min\{q_1, r\}$.
\end{theorem}


Theorem \ref{th-1} quantifies how the convergence rate of the empirical bandwidth is affected by the choice of the pilot bandwidth and order $h_1$ and $q_1$, and the rate of decay of the autocovariance functions in norm. The optimal rate of convergence is achieved by taking $\kappa=1/(2q + 2\alpha +1)$, which depends on the unknown rate of decay of the autocovariance functions. In the ``worst case scenario", when $q =r =1$, then taking $h_1=AT^{-1/5}$ achieves a rate of approximation on the order of $T^{-1/5}$ for the estimated optimal bandwidth, which compares to the results of \cite{buhlmann:1996} in the scalar case. This would be the rate achieved by following the advice of \cite{panaretos:tavakoli:2012} to choose pilot estimates.

It also follows from Theorem \ref{th-1} that $q_1$ is a potential limiting factor for the convergence rate of the empirical bandwidth; if it is always chosen to be larger than $r$ in Assumption \ref{as-r}, then the rate of approximation to the optimal bandwidth can achieve the order of $T^{-r/(2q + 2r +1)}$. This observation that was made by Politis and Romano \cite{politis:romano:1996}. In particular, if $r$ in Assumption \ref{as-r} may be taken to be arbitrarily large with $\kappa=1/r$, then the rate of approximation approaches the parametric rate of $T^{-1/2}$. The fact that the order of the pilot weight function may limit the convergence rate of the estimated bandwidth motivates the idea of choosing the pilot weight function to be in the family of infinite order ``flat-top" weight functions, which has been thoroughly explored in \cite{politis:romano:1996,politis:romano:1999,politis:2003adap}. A flat top weight function $W_\infty$ is of the form
\beq\label{flattop}
W_\infty(t)=\left\{
\begin{array}{ll}
1,\quad &0\leq |t| <k_1\\
\frac{k_2-|t|}{k_2-k_1},\quad &k_1 \leq |t| <k_2\\
0,\quad &|t|\geq k_2
\end{array}
\right.
\eeq
where $k_2 > k_1$. In the case when the autocovariance kernels decay quickly, then using the flat top kernel for pilot estimation can substantially increase the rate of approximation of the asymptotically optimal bandwidth. Below, we suppose the rate of decay of the autocovariance kernels satisfies the following condition.

\begin{assumption}\label{as-e}
There exist positive constants $D$ and $d$ such that $\|\gamma_\ell\| \le De^{-d|\ell|}$.
\end{assumption}
Assumption \ref{as-e} is satisfied by many functional time series of interest. For example, it holds when the functional time series $X_t(u)$ satisfies a FAR model of order one in which the autoregressive operator has norm less than one; see Lemma 3.1 in \cite{bosq:2000}. It also clearly holds for observations with a bounded range of dependence.

\begin{theorem}\label{th-2}
If Assumption \ref{as-e} holds, and $h_1(T)= A \log(T)$ for some constant $A>1/2dk_1$, where $k_1$ is defined in \eqref{flattop} and Assumption \ref{as-e}, then
\begin{equation*}
\hat{h}_{opt}(h_1,\infty) = h_{opt}\left(1+O_P\left(\frac{\log^{(2q+1)/2}(T)}{\sqrt{T}}\right)\right).
\end{equation*}
\end{theorem}

We explore the use of the flat-top weight function as well as other popular weight function choices to obtain pilot estimates in the simulation study below.

\section{Simulation Study}\label{simul}

We utilize Monte Carlo methods in order to evaluate the performance of the ``plug-in" bandwidth selection procedure introduced above. The ultimate goal of our simulation study is to provide practical, empirically motivated advice on how to choose the bandwidth parameter $h$ in order to minimize $\|\hat{C}_{h,q} - C\|^2$ when using a finite order weight function. To this end, we consider a number of different weight functions $W_q$ and data generating processes (DGPs) with varying structures and levels of interdependence.

\subsection{Outline}

The weight functions that we considered for computing the covariance estimators are:
\begin{align}
W_{\text{BT}}&= \left\{ \begin{array}{ll}
         1-|x| & \mbox{for $|x| \leq 1$};\\
        0 & \mbox{otherwise}.\end{array} \right. \tag{Bartlett} \\
W_{\text{PR}} &= \left\{ \begin{array}{ll}
 1-6x^2 + 6|x|^3 & \mbox{for $0\leq |x|\leq \frac{1}{2}$}; \\
 2(1-|x|)^3 & \mbox{for $\frac{1}{2}\leq |x|\leq 1$}; \\
 0 & \mbox{otherwise}. \end{array} \right. \tag{Parzen} \\
 W_{\text{TH}} &= \left\{ \begin{array}{ll}
 (1+\cos(\pi x))/2 & \mbox{for $|x|\leq 1$}; \\
 0 & \mbox{otherwise}. \end{array} \right. \tag{Tukey-Hanning} \\
  W_{\text{QS}} &= \frac{25}{12\pi^2 x^2}\left(\frac{\sin(6\pi x/5)}{6\pi x/5} - \cos(6\pi x/5)\right) \tag{Quadratic spectral}
\end{align}
These weight functions have also been considered in \cite{andrews:1991} and \cite{andrews:monahan:1992}, and their corresponding orders are $1,2,2,$ and $2$, respective to the above list. In order to define the DGPs that we considered, let  $\{W_i(t), -\infty<i<\infty, t\in [0,1]\}$ denote independent and identically distributed standard Brownian motions. We generated data according to:
\begin{align*}
\text{MA}_{\psi}(p):& \quad X_i(t) = W_i(t) + \sum^p_{j=1}\int \psi(t,s) W_{i-j}(s)ds \\
\text{FAR}_{\psi}(1):& \quad  X_i(t) = \int \psi(t,s)X_{i-1}(s)ds + W_i(t) \\
\text{MA}_{\phi}^*(p):& \quad X_i(t) = W_i(t) + \phi \sum^p_{j=1}W_{i-j}(t) \\
\text{FAR}_{\phi}^*(1):& \quad X_i(t) = \phi X_{i-1}(t) + W_i(t)
\end{align*}
We considered the processes MA$_1^*(0)$, MA$_{0.5}^*(1)$, MA$_{0.5}(4)$, MA$_{\psi_1}(4)$, MA$_{0.5}^*(8)$, FAR$_{0.5}^*(1)$ and FAR$_{\psi_2}(1)$, where $\psi_1(t,s) = 0.34\exp^{\frac{1}{2}(t^2+s^2)}$ and $\psi_2(t,s) = \frac{3}{2}\min(t,s)$. The choice of the constants in the definition of $\psi_1$ and $\psi_2$ is done so that $\|\psi_1\|\approx \|\psi_2\|\approx 0.5$.

\begin{figure}[!htbp]
\centering
\includegraphics[width=7.8cm]{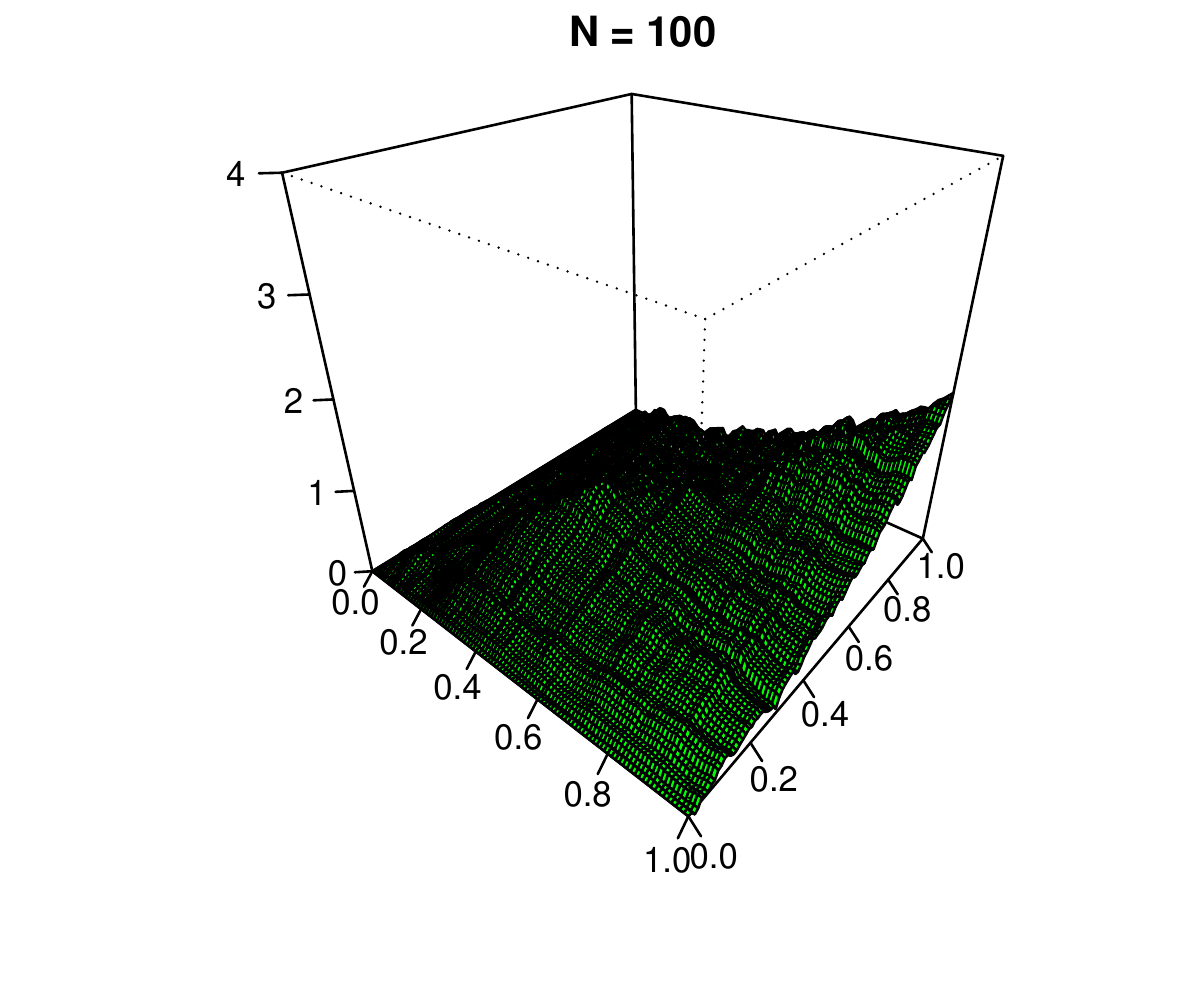}
\includegraphics[width=7.8cm]{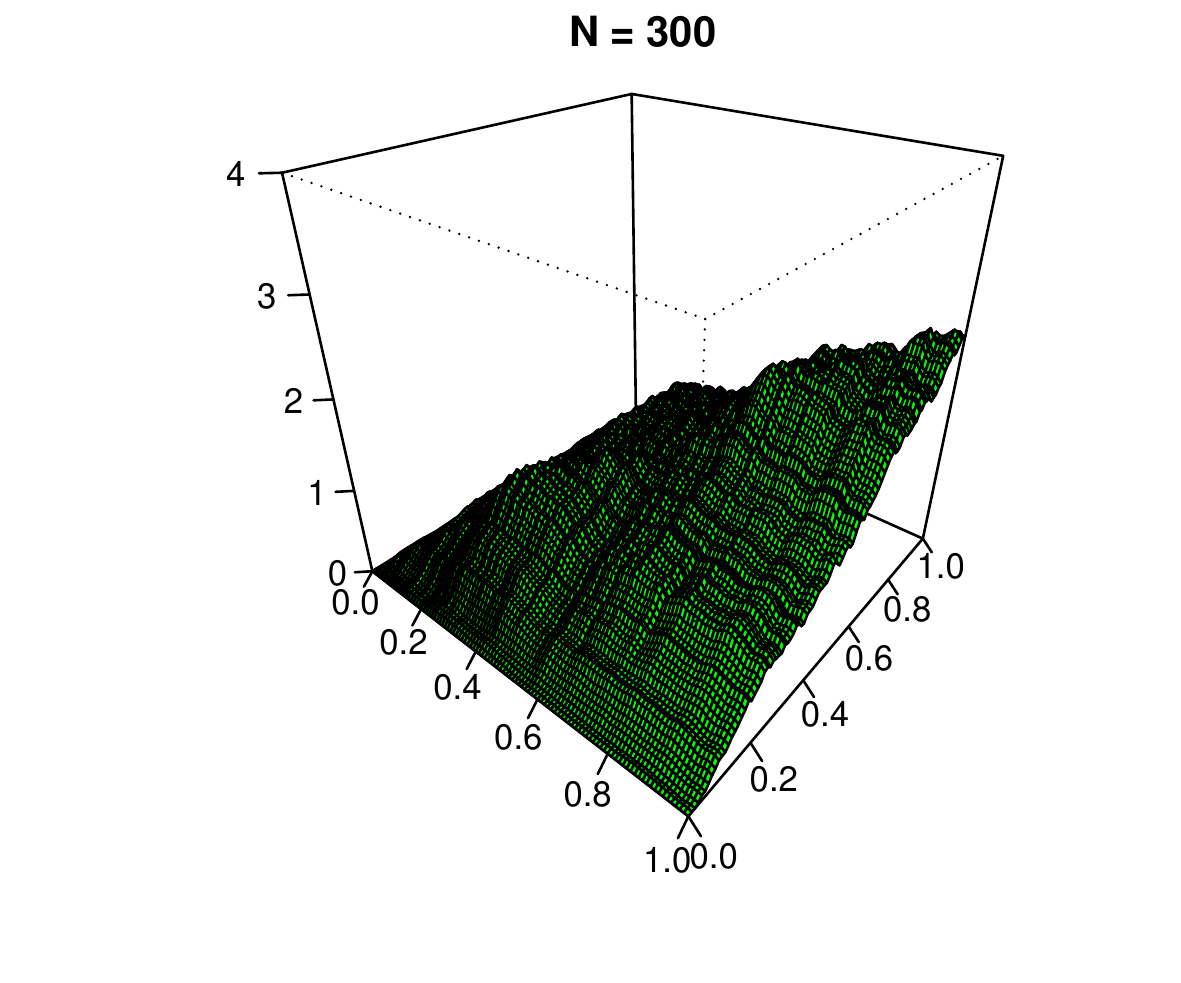}
\\
\includegraphics[width=7.8cm]{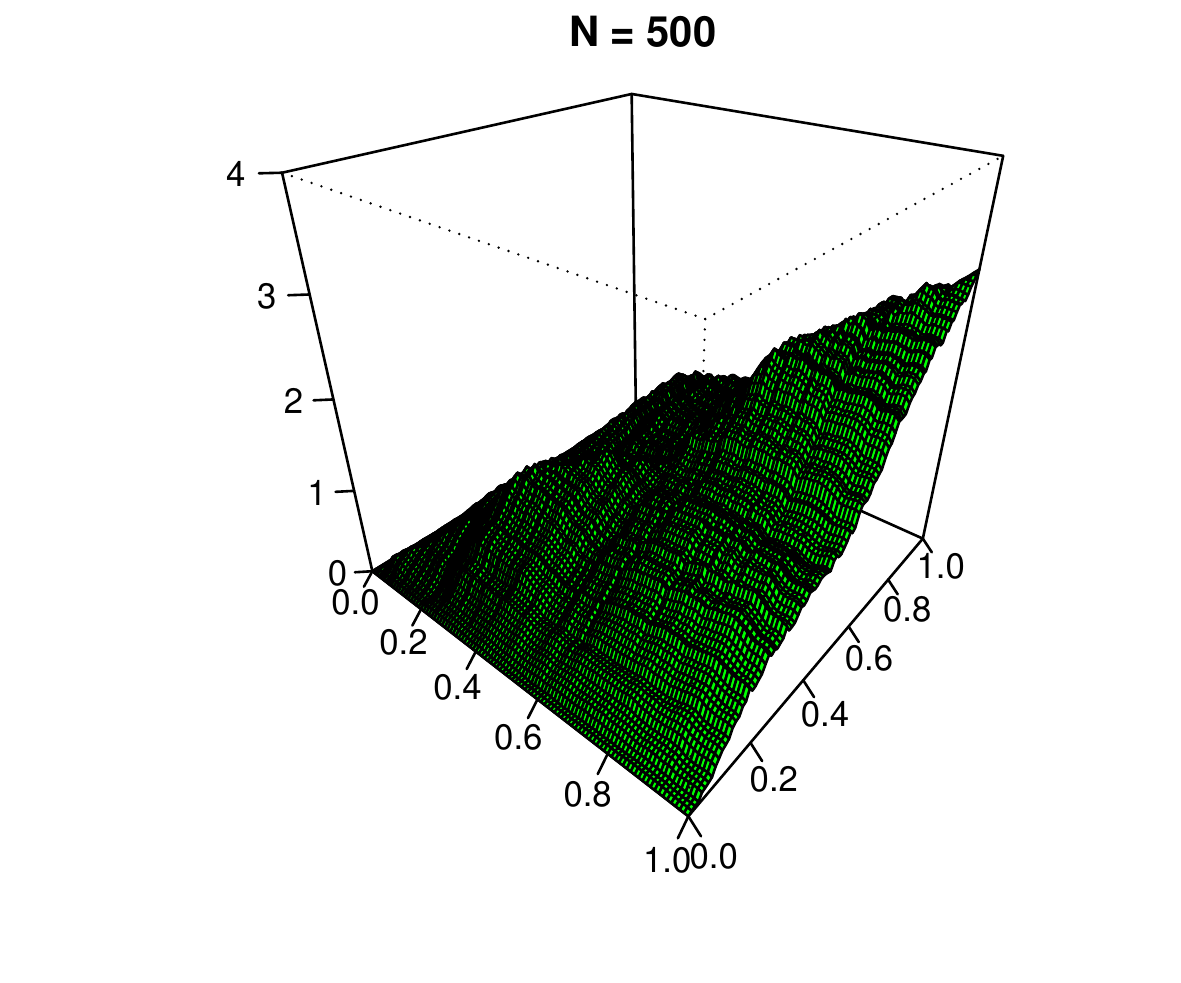}
\includegraphics[width=7.8cm]{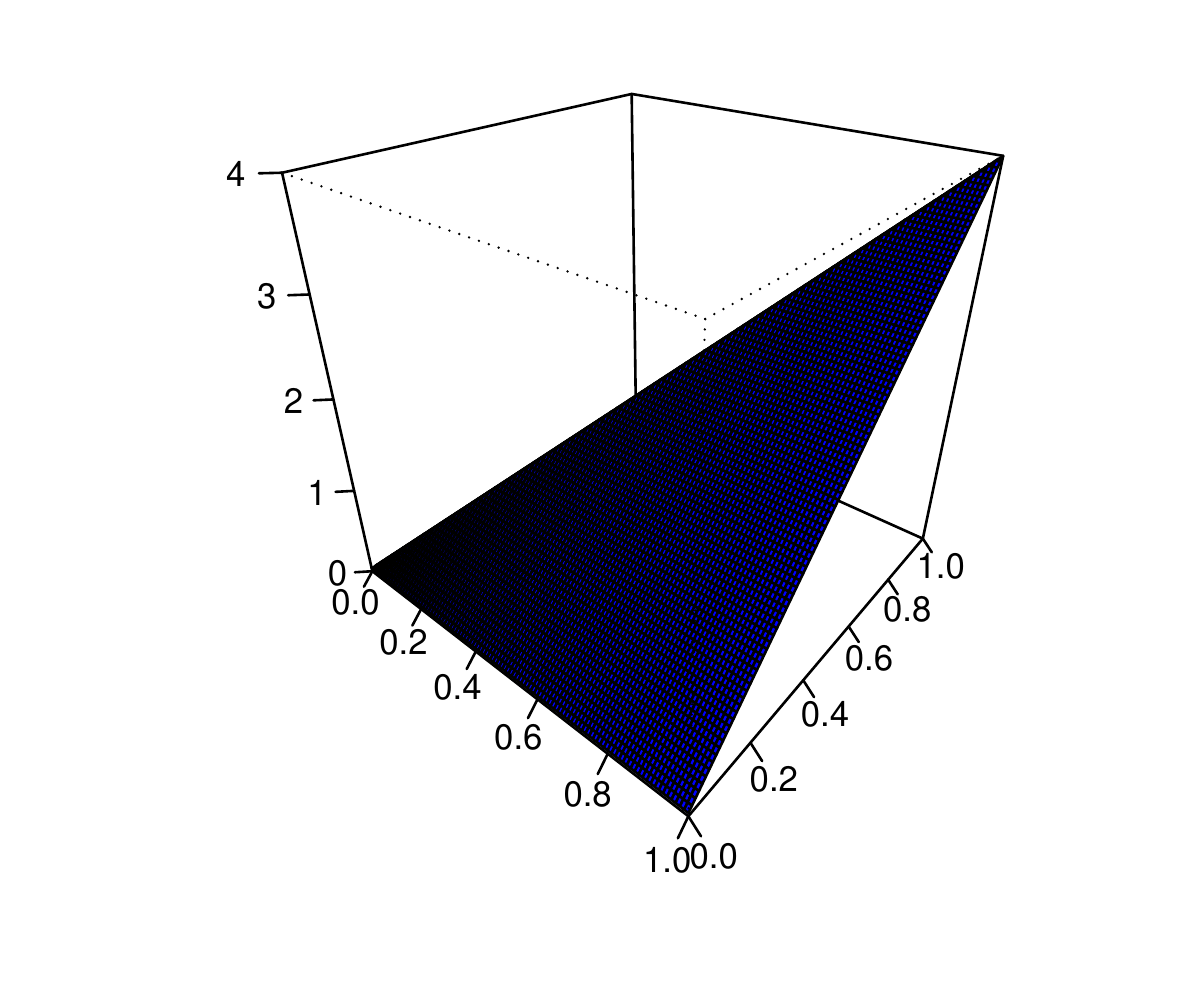}
\caption{Lattice plots of the long run covariance weight function estimators with FAR$^*_{0.5}(1)$ data using the Bartlett kernel with the proposed plug-in bandwidth for values of $T=100, 300$ and 500 along with the theoretical long run covariance (lower right).}\label{fig:1}
\end{figure}

As an illustration of the estimators $\hat{C}_{h,q}$, Figure~\ref{fig:1} shows lattice plots of the long-run covariance function estimators with FAR$_{0.5}^*(1)$ data, using the plug-in bandwidth selection procedure for $T=100, 300$ and 500, as well as the theoretical long-run covariance.

When the kernels $\psi_1$ and $\psi_2$ are used to define a DGP, then it is not tractable to compute $C$ explicitly. In these cases, $C$ is replaced by the approximation
\begin{equation}
C^*(t,s) = \sum^{10^4}_{j=1}\bar{X}_j(t)\bar{X}_j(s),
\end{equation}
where
\begin{equation}
\bar{X}_j(t) = \frac{1}{10^4}\sum^{10^4}_{i=1}X_i^{(j)}(t)
\end{equation}
and $X_i^{(j)}(t)$'s are computed according to data generating process MA$_{\psi_1}(4)$ or FAR$_{\psi_2}(1)$ independently for each $j$. The approximation is reasonably accurate, since $C$ is the limiting covariance of $\sqrt{T}\bar{X}(u)$. We considered sample sizes of $T=100,300,$ and $500$.

For each finite order weight function $W_q$ described above, we compared five settings in order to select the bandwidth parameter $h$:

\begin{enumerate}
\item[Setting 1:] $h=T^{1/5}$
\item[Setting 2:] $h=T^{1/4}$
\item[Setting 3:] $h$ is selected according to the proposed method with $W_{q_1}=W_{q}$, and $h_1=T^{1/5}$.
\item[Setting 4:] $h$ is selected according to the proposed method with $W_{q_1}=W_{\infty}$, with $k_1=1/2$, $k_2=1$, and $h_1=T^{1/5}$.
\item[Setting 5:] $h$ is selected according to the proposed method with $W_{q_1}=W_{\infty}$, with $k_1=1/2$, $k_2=1$, and $h_1$ is chosen according to the adaptive method of Horv\'ath et al \cite{horvath:rice:whipple:2014}.
\end{enumerate}

Namely in Settings 1 and 2, the bandwidth is just a fixed function of the sample size, while Settings 3,4 and 5 compare the proposed method over several different choices for the initial weight function and bandwidth.

\subsection{Results}

\begin{figure}[!htbp]
\centering
\includegraphics[width=7.62cm]{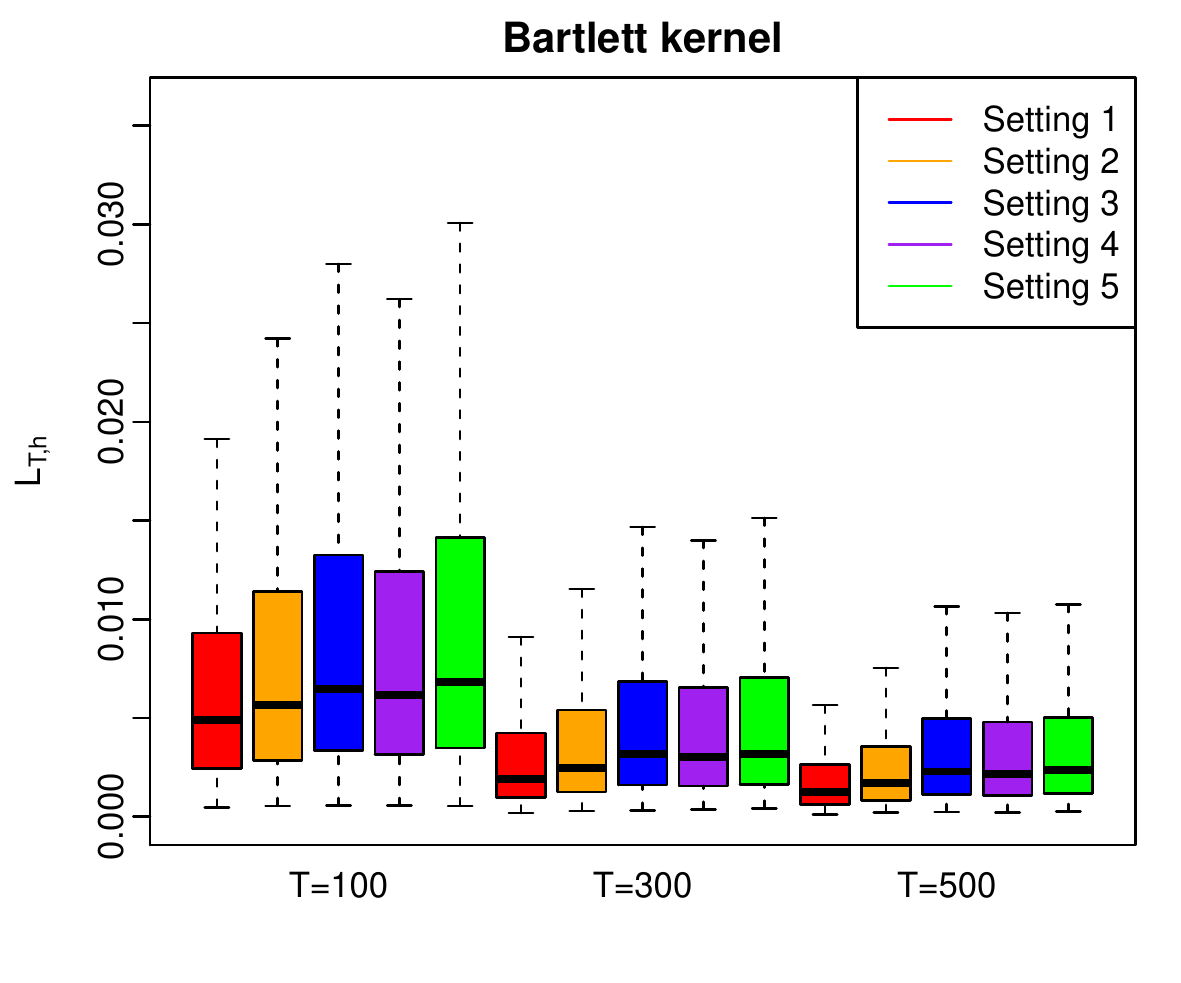}
\quad
\includegraphics[width=7.62cm]{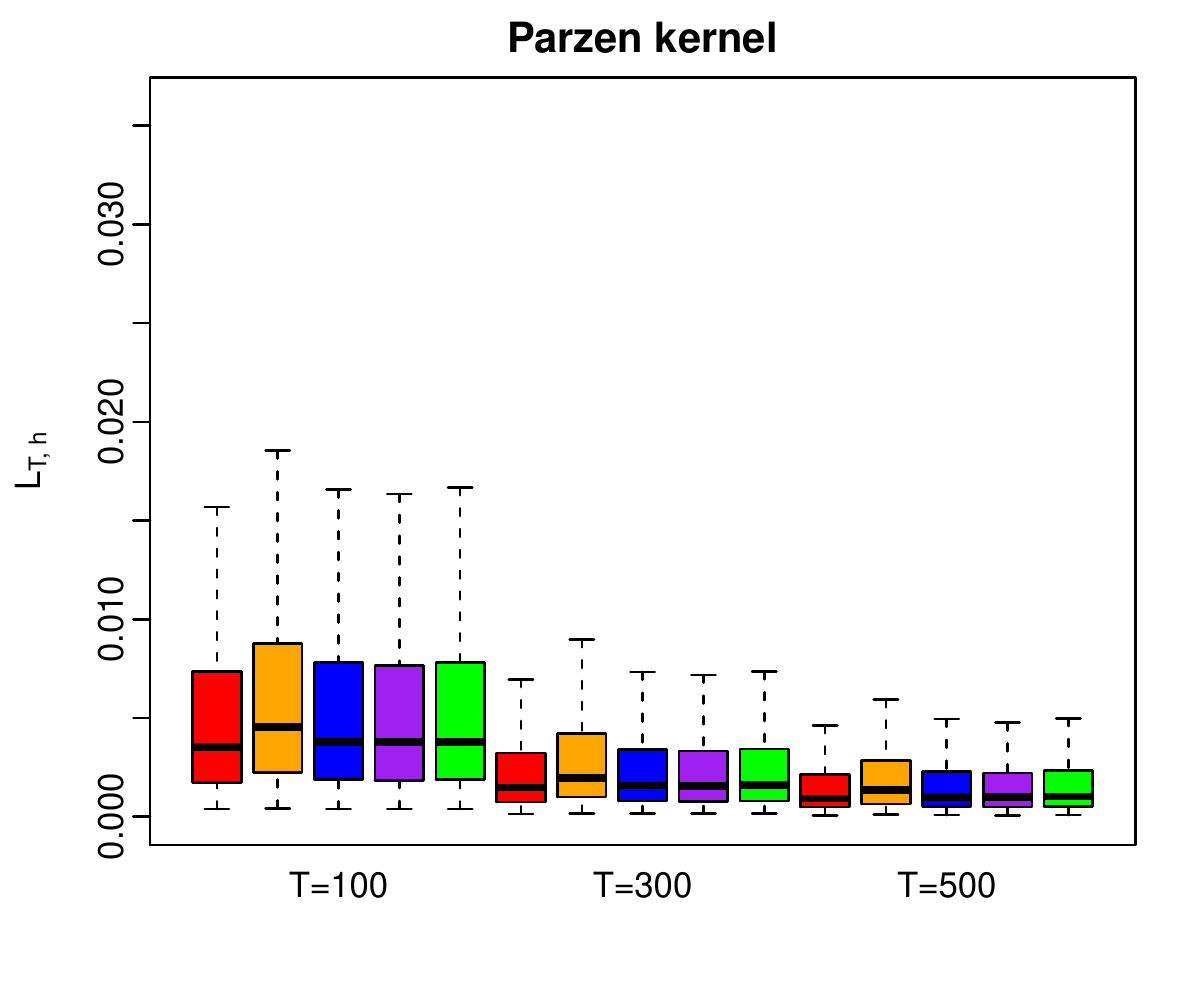}
\\
\includegraphics[width=7.62cm]{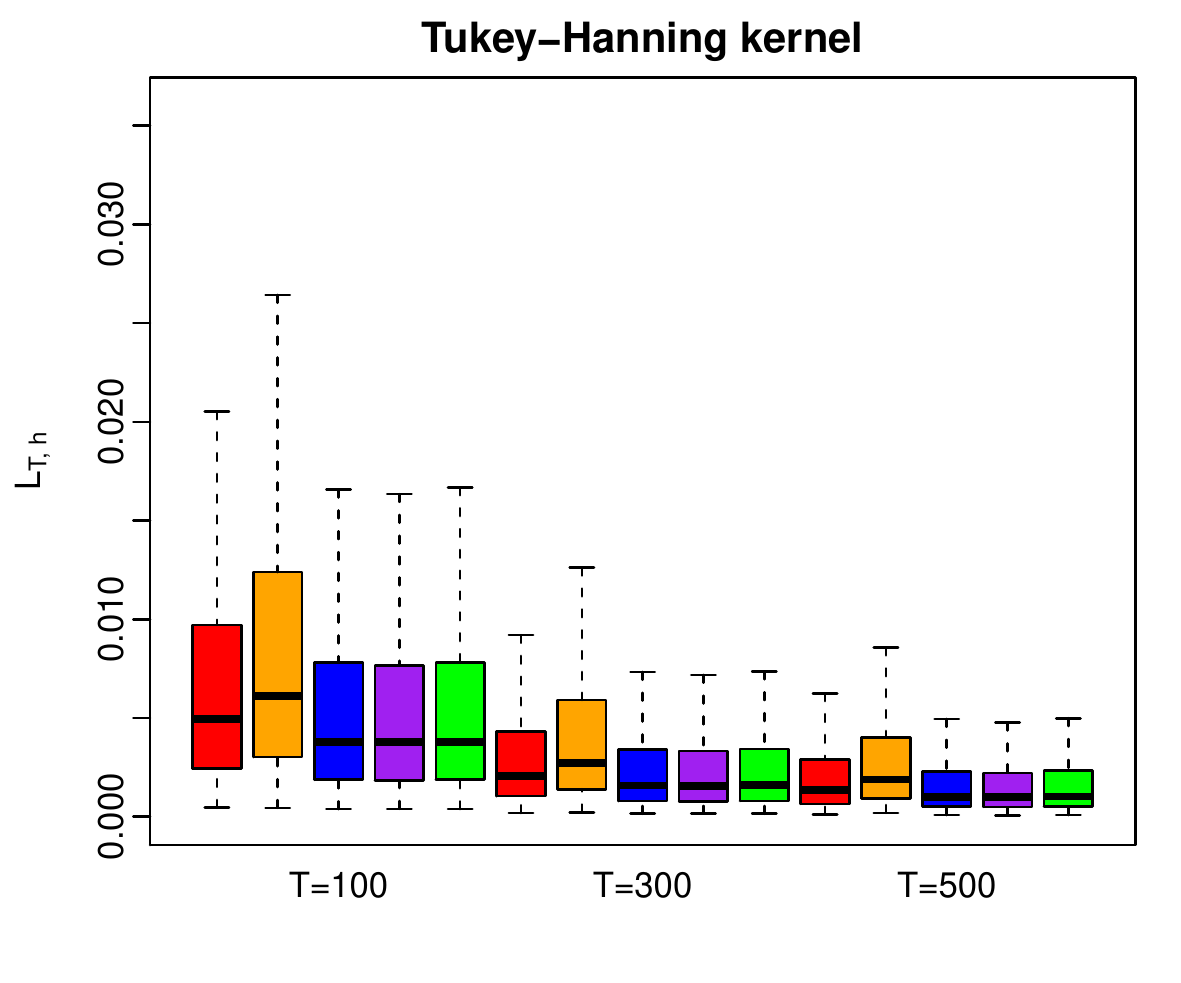}
\quad
\includegraphics[width=7.62cm]{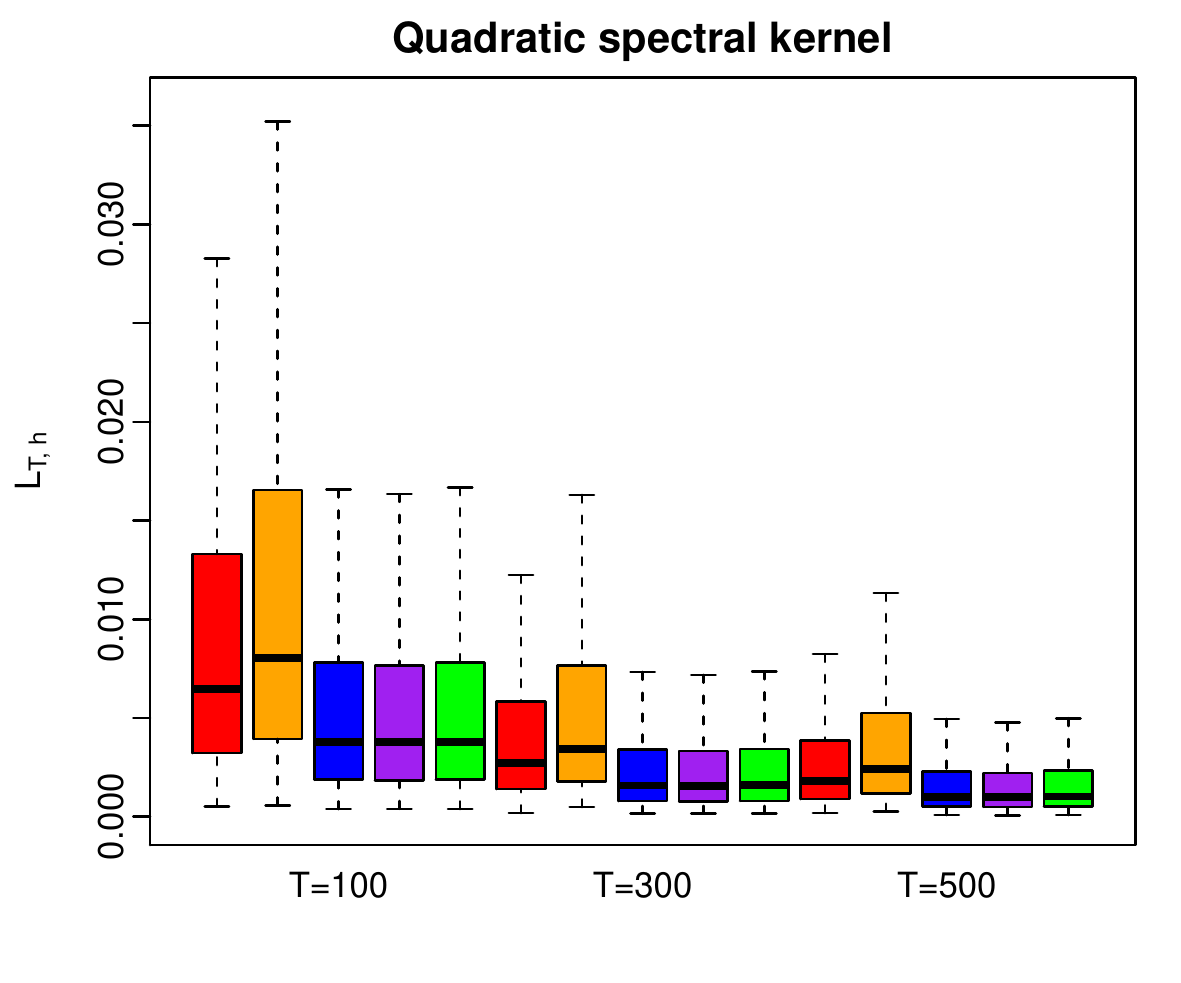}
\caption{Results for MA$_1(0)$ with estimated bandwidths.}\label{box-1}
\end{figure}

\begin{figure}[!htbp]
\centering
\includegraphics[width=7.62cm]{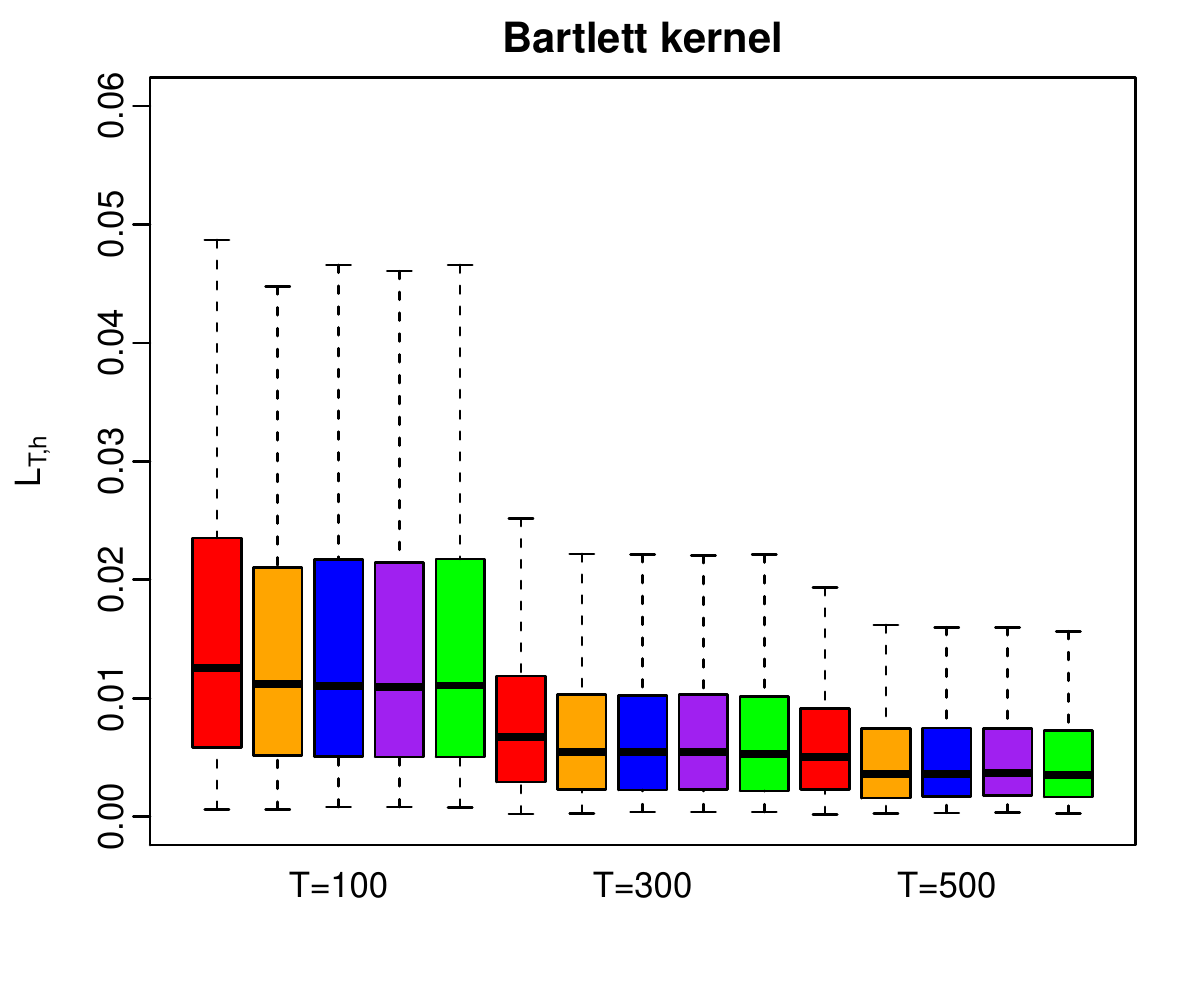}
\quad
\includegraphics[width=7.62cm]{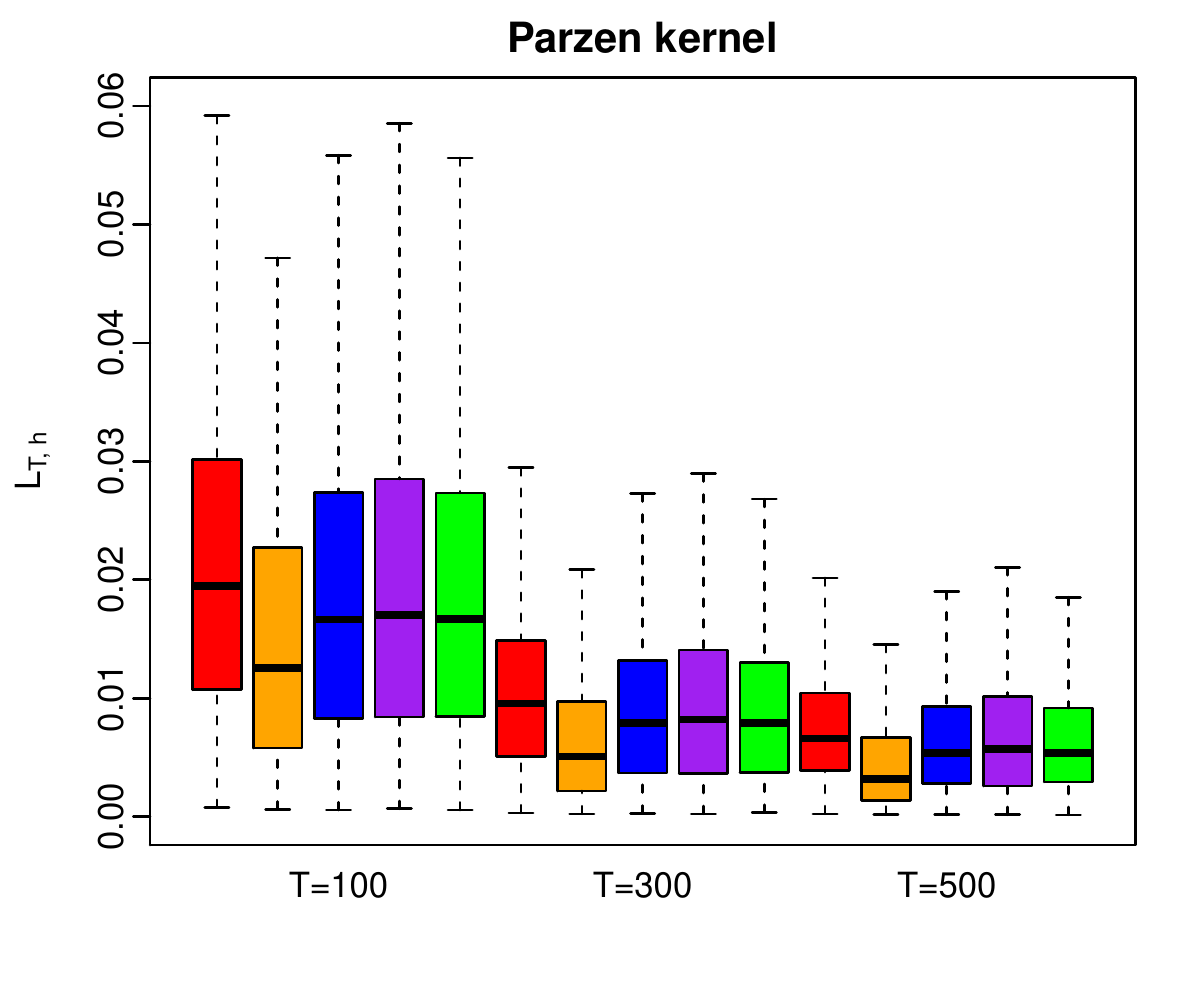}
\\
\includegraphics[width=7.62cm]{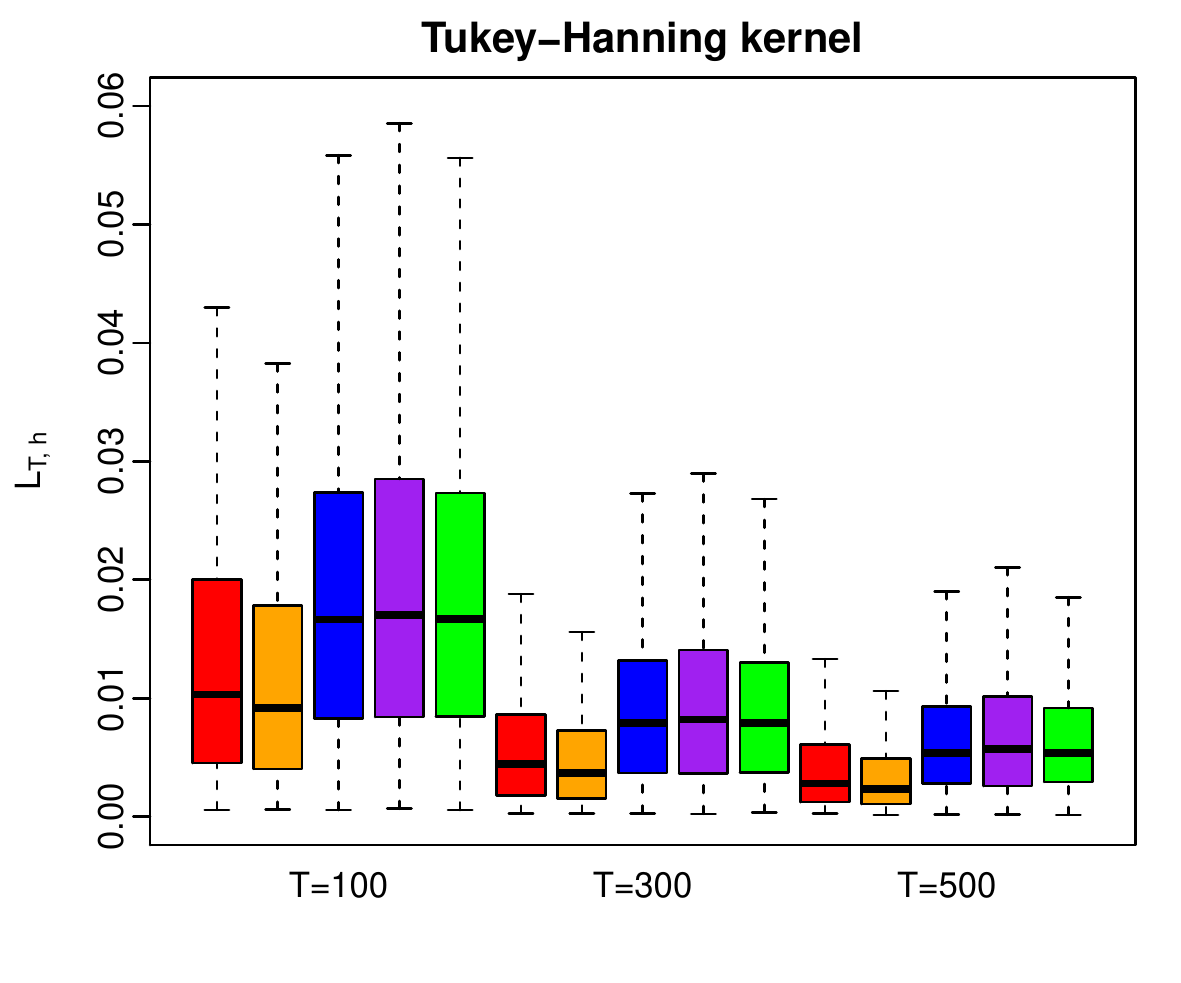}
\quad
\includegraphics[width=7.62cm]{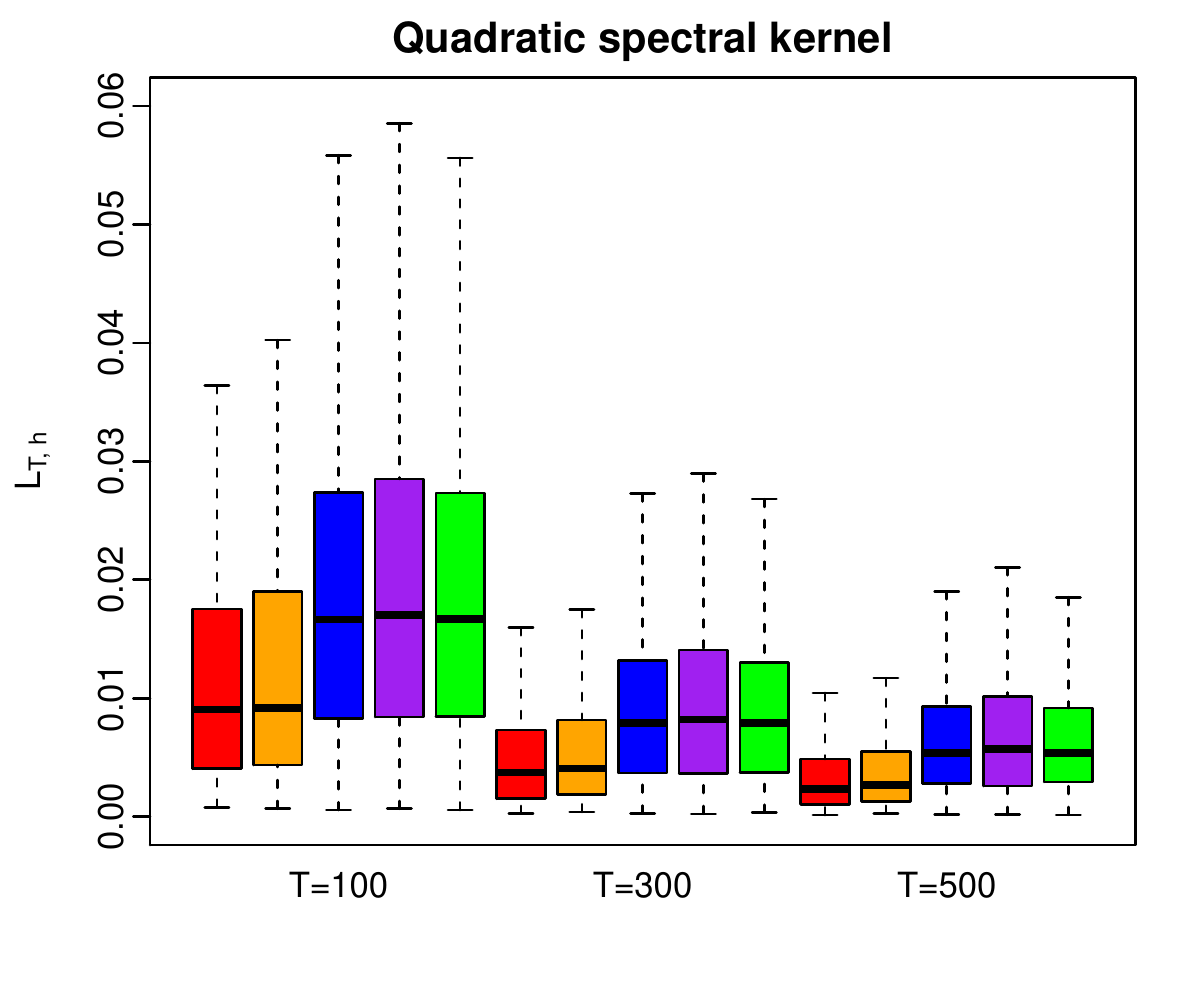}
\caption{Results for MA$_{0.5}(1)$ with estimated bandwidths.}\label{box-2}
\end{figure}

\begin{figure}[!htbp]
\centering
\includegraphics[width=7.62cm]{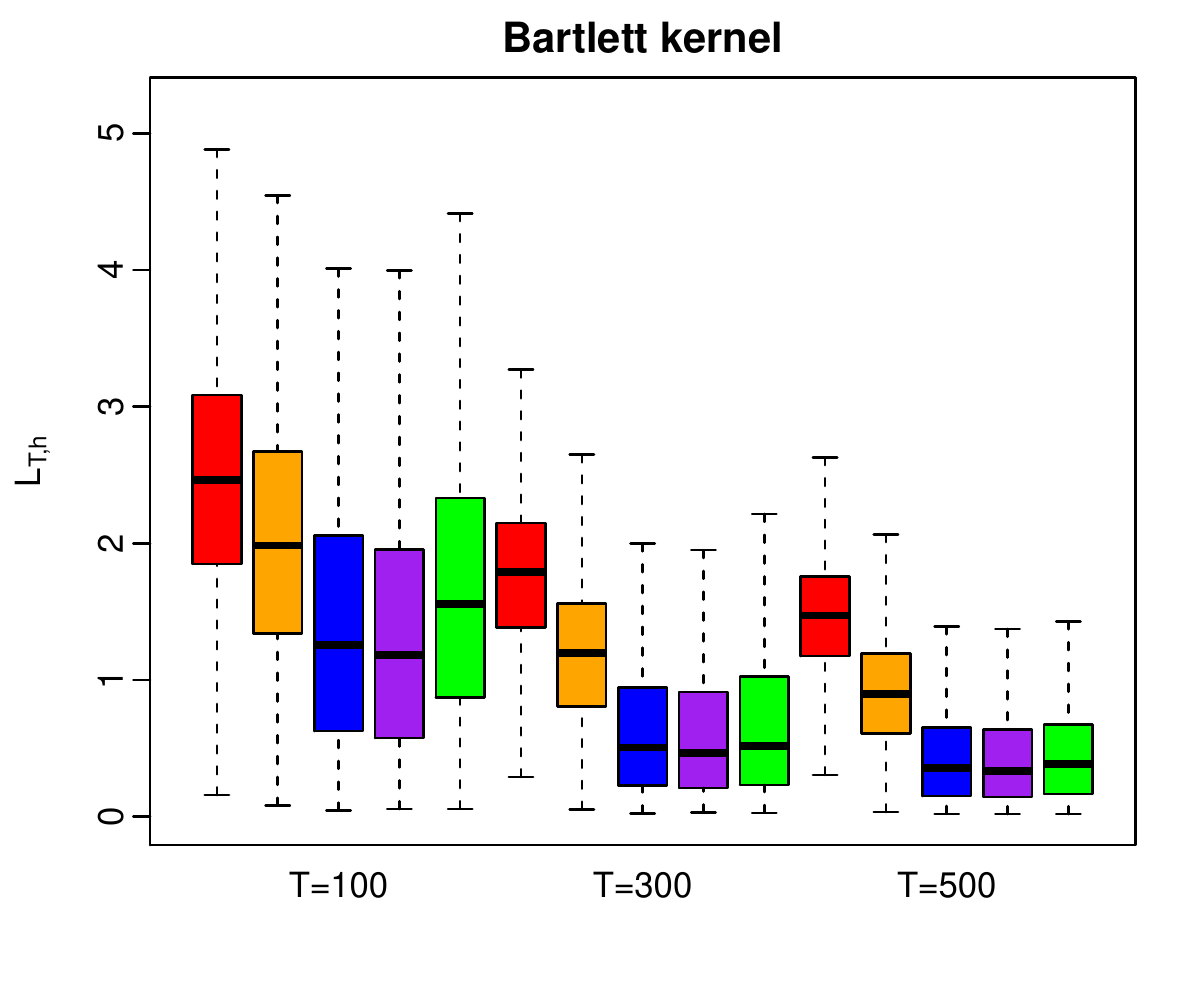}
\quad
\includegraphics[width=7.62cm]{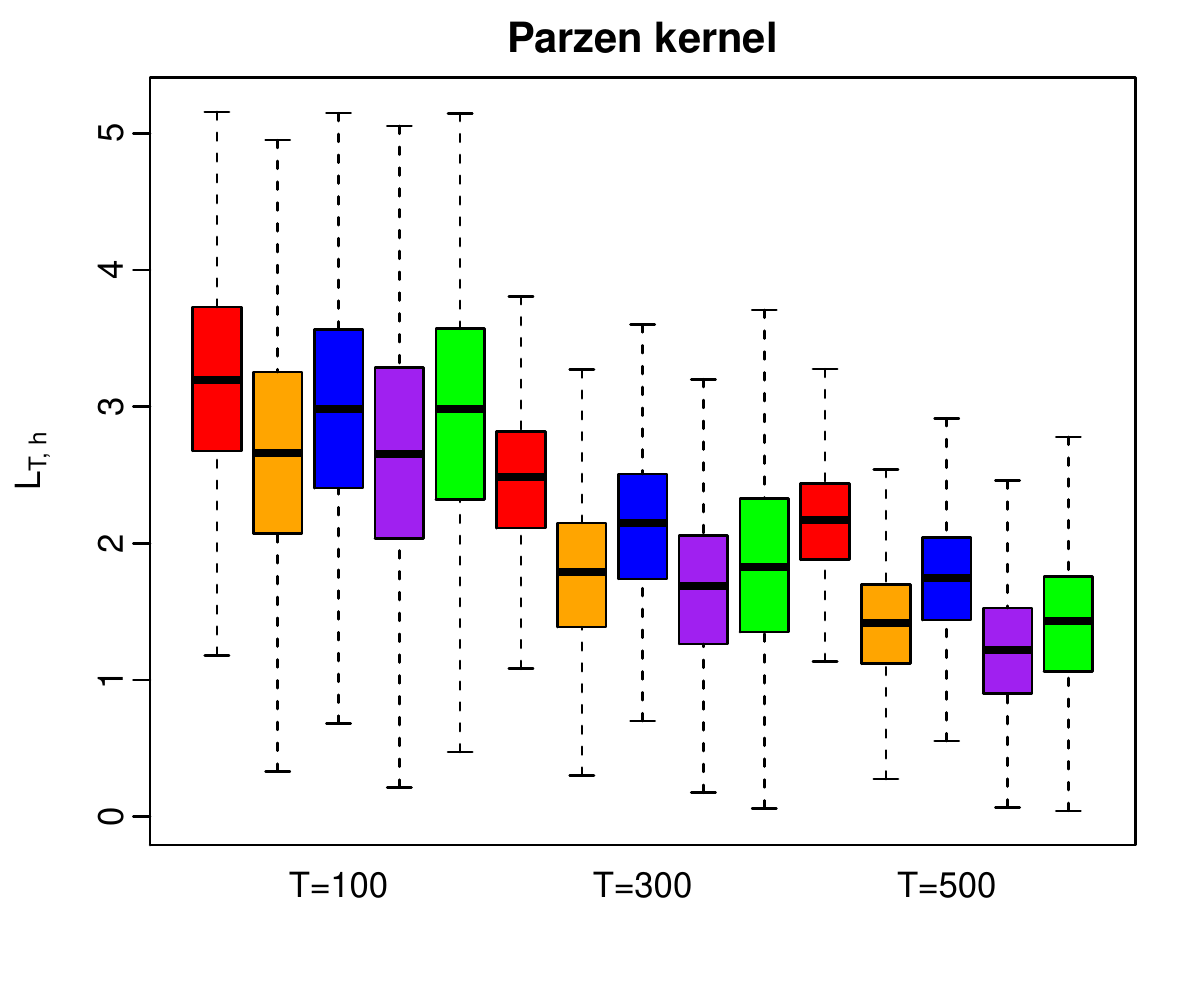}
\\
\includegraphics[width=7.62cm]{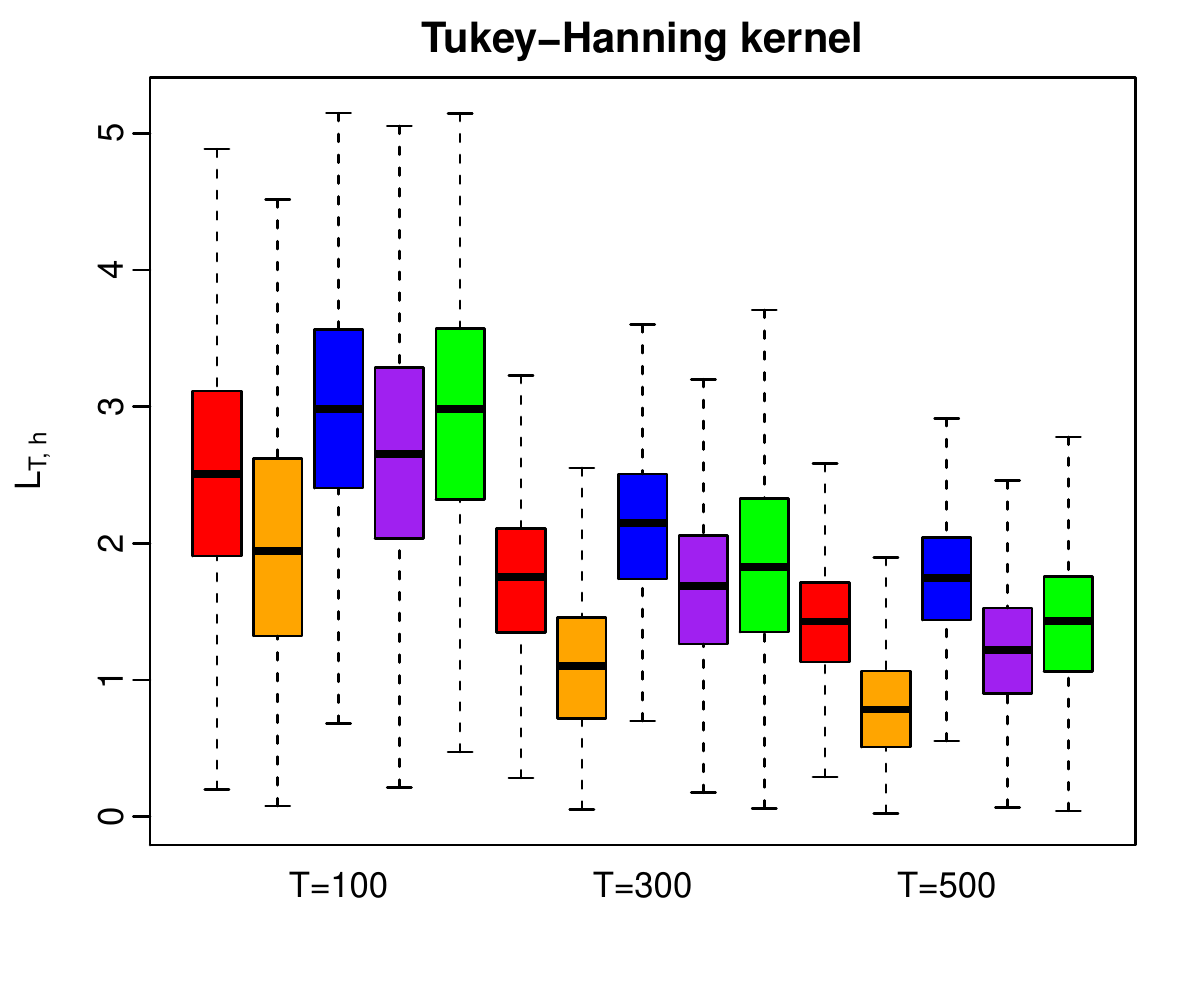}
\quad
\includegraphics[width=7.62cm]{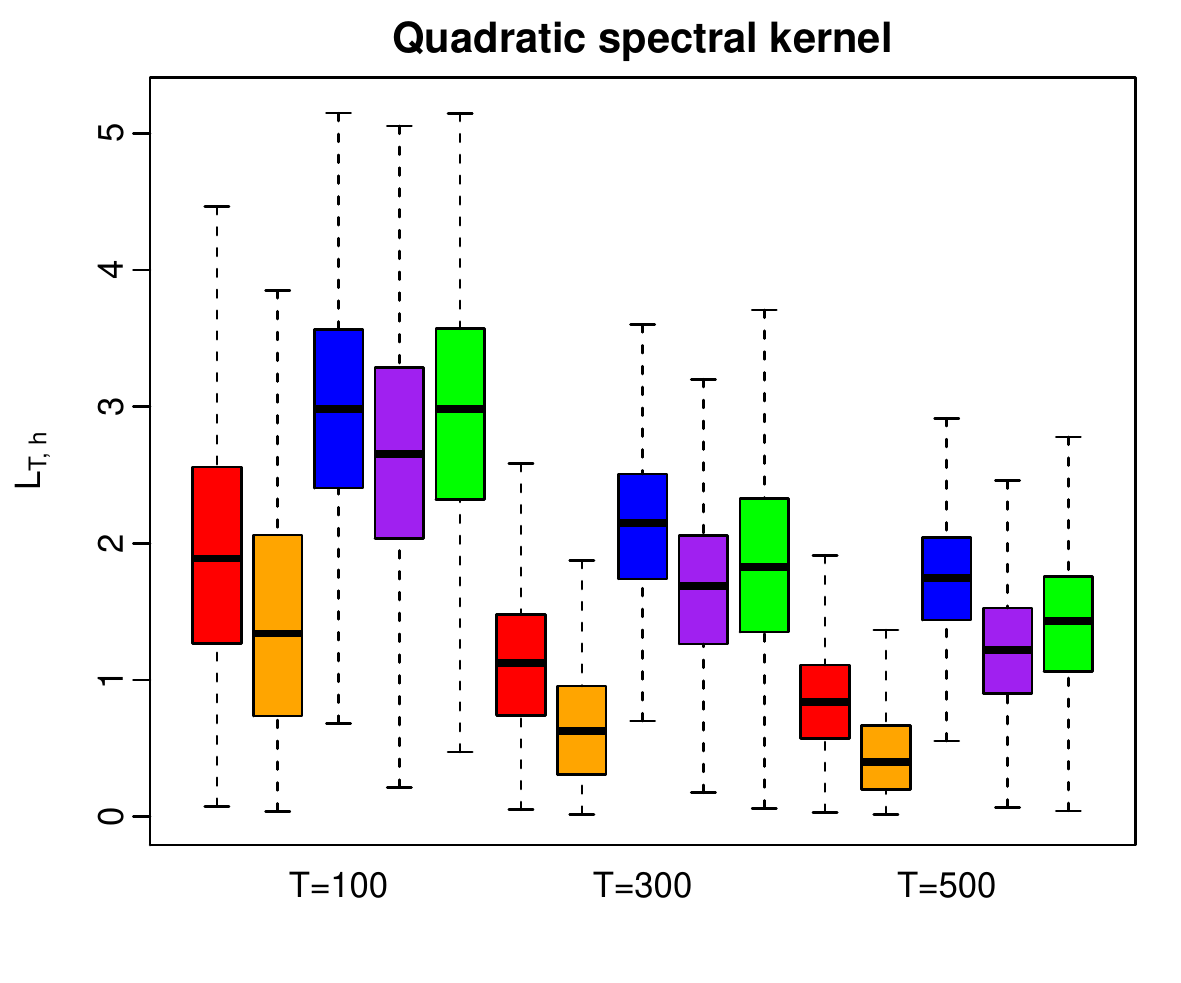}
\caption{Results for MA$_{0.5}(4)$ with estimated bandwidths.}\label{box-3}
\end{figure}

\begin{figure}[!htbp]
\centering
\includegraphics[width=7.62cm]{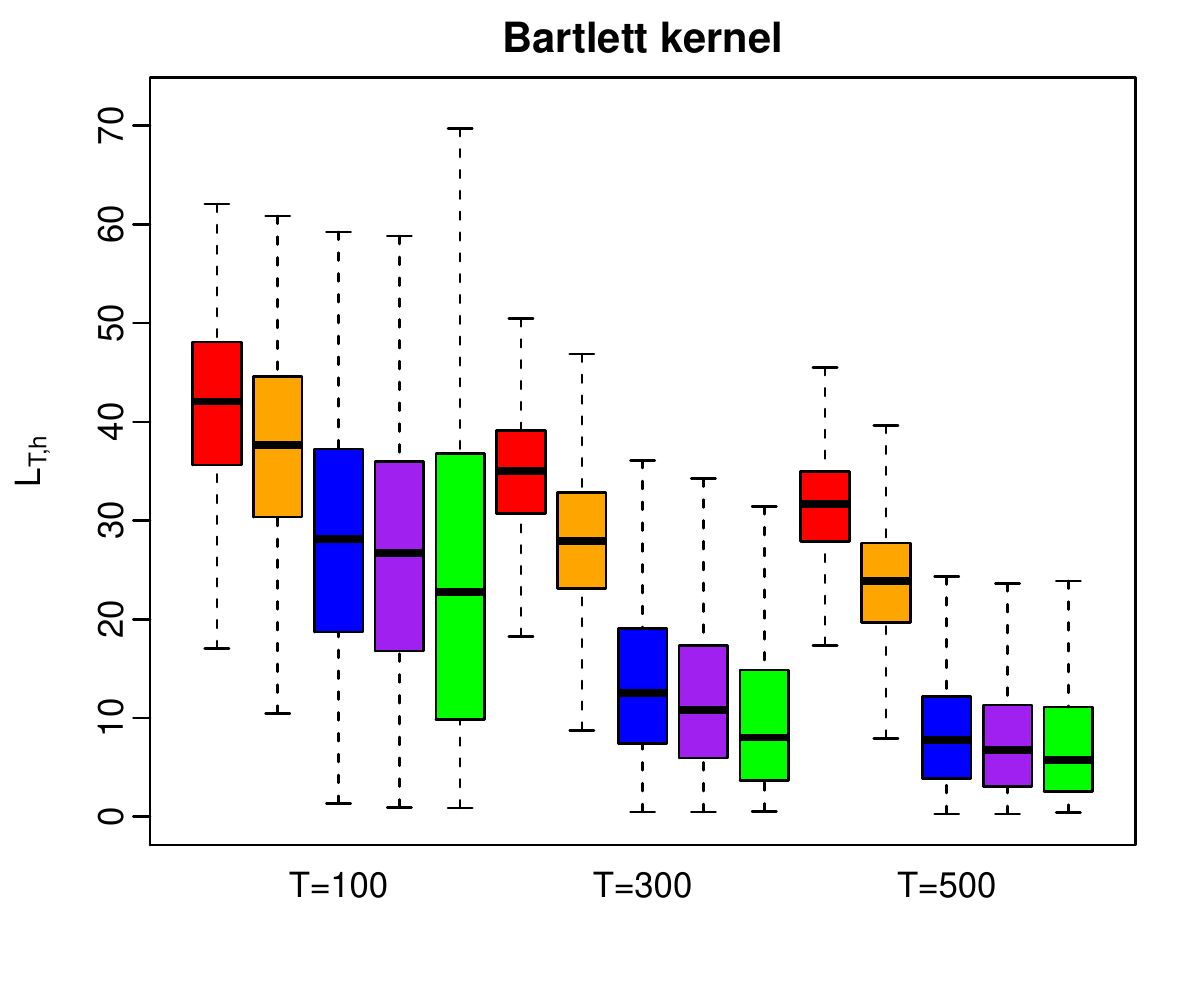}
\quad
\includegraphics[width=7.62cm]{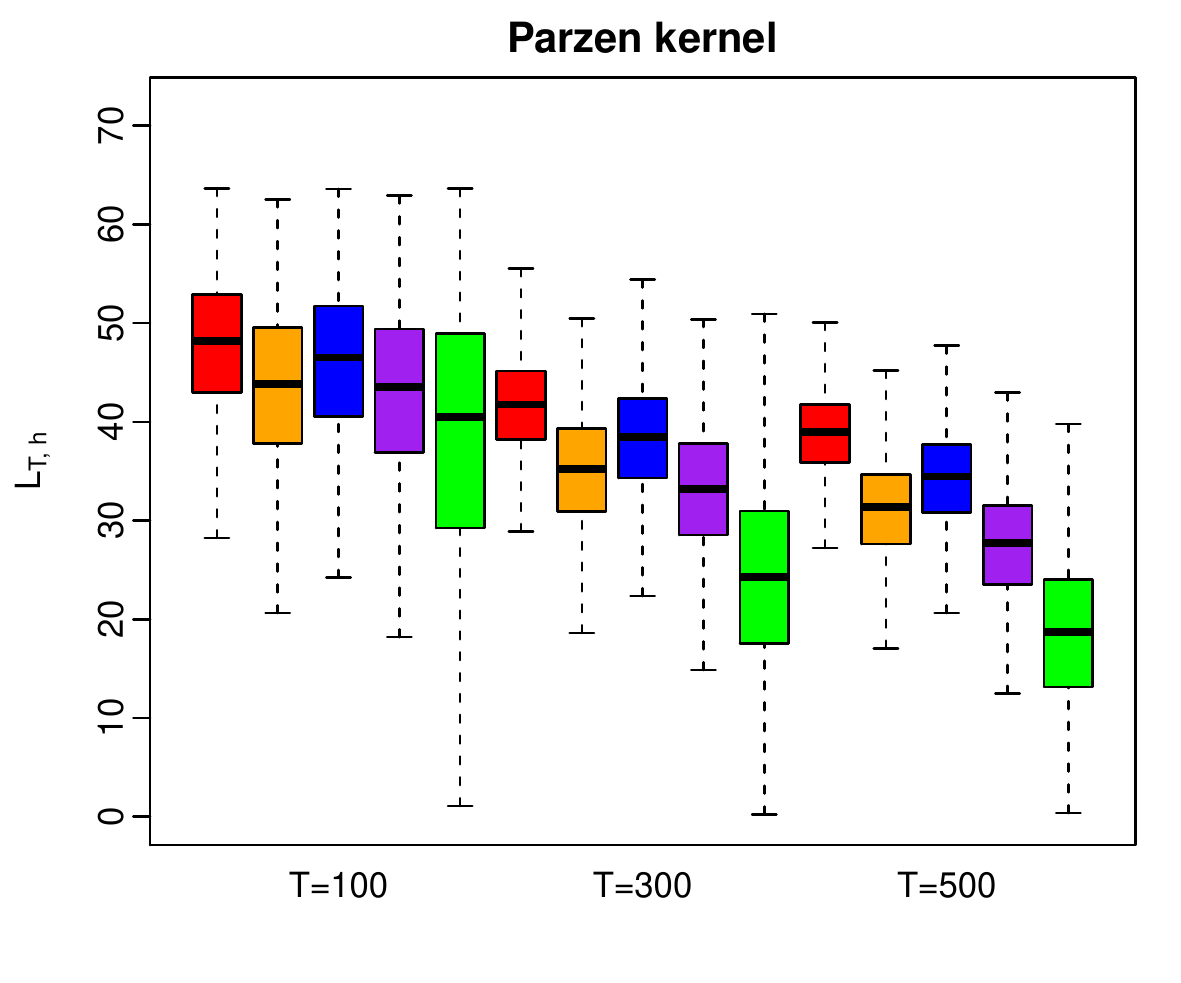}
\\
\includegraphics[width=7.62cm]{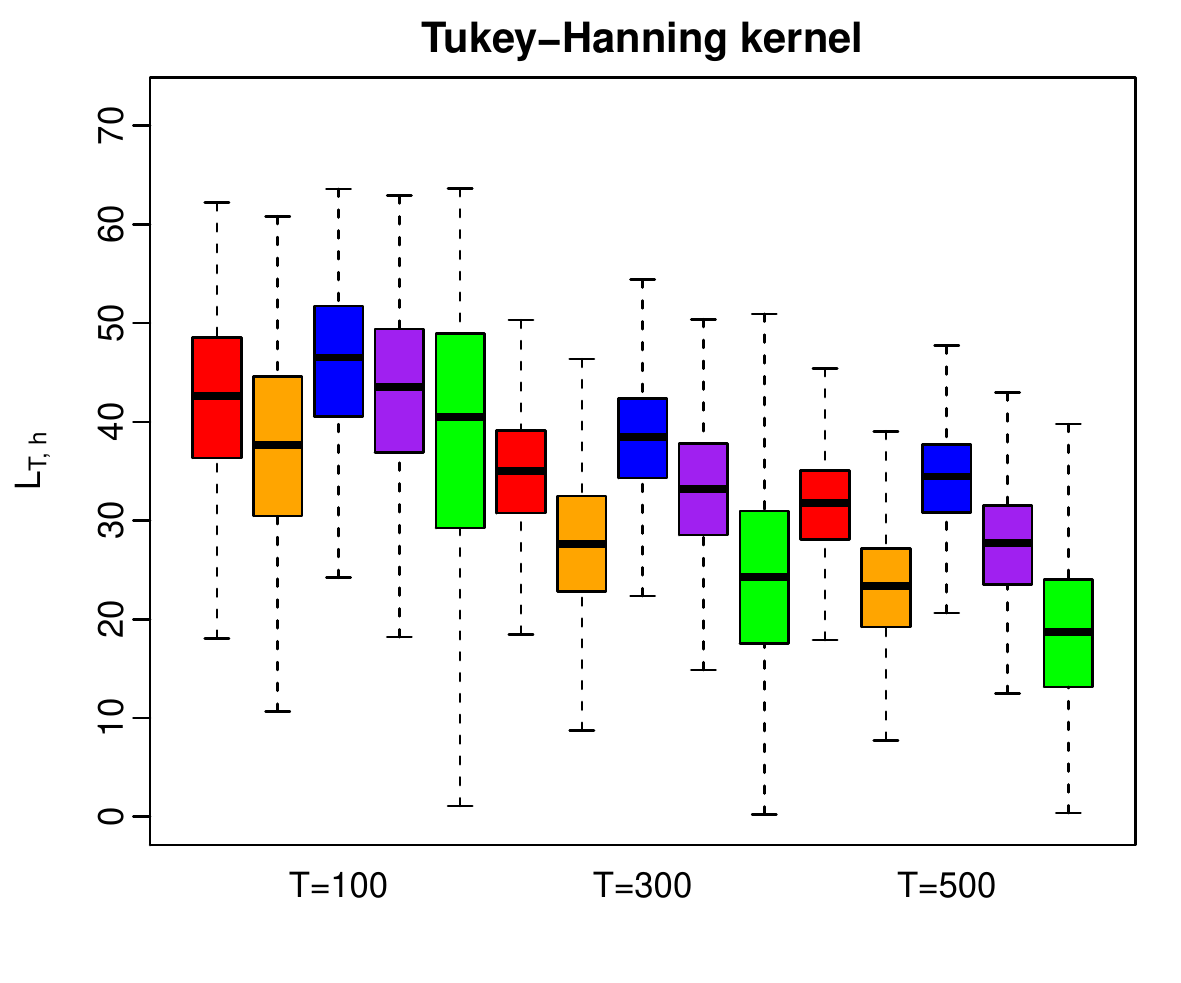}
\quad
\includegraphics[width=7.62cm]{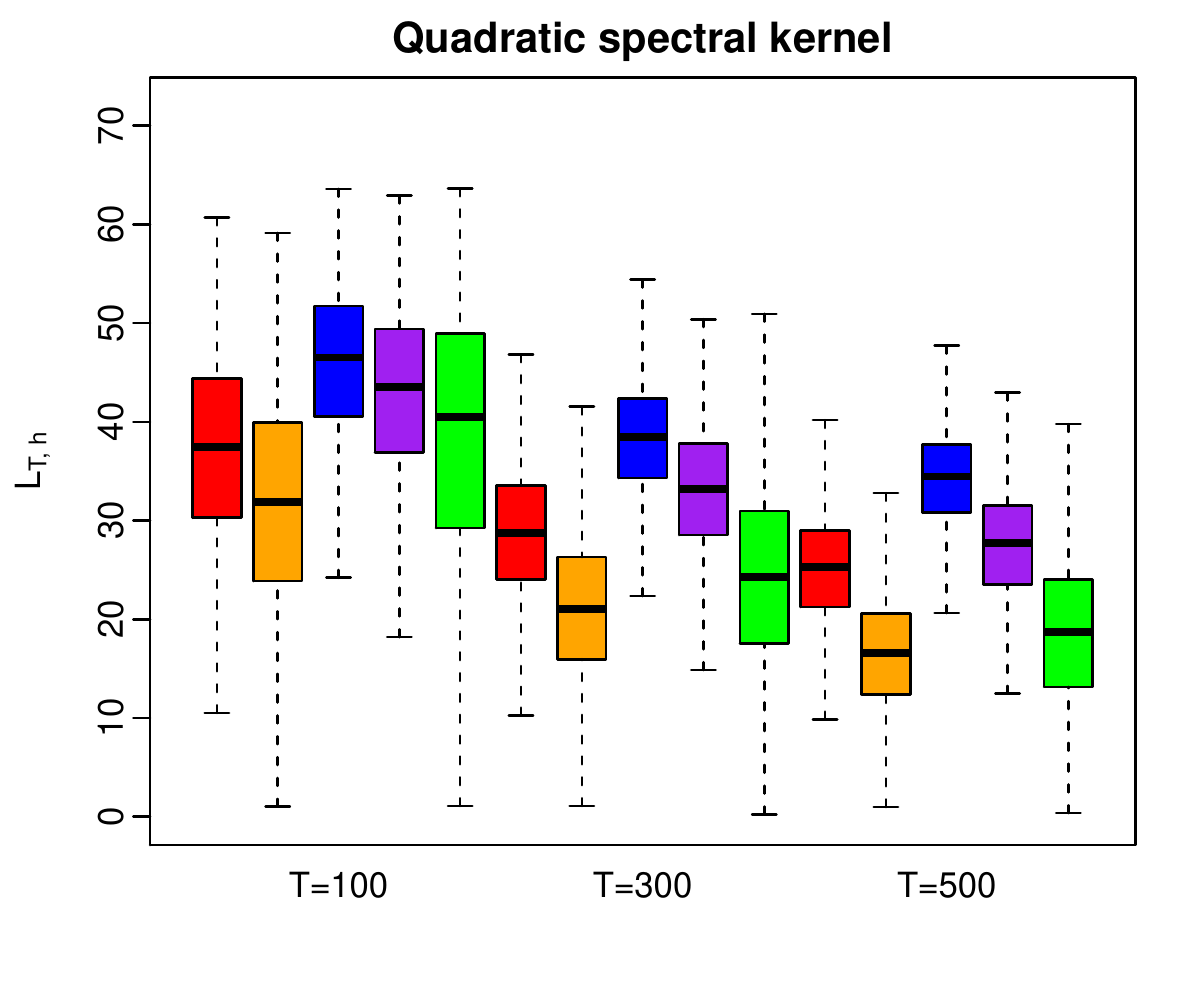}
\caption{Results for MA$_{0.5}(8)$ with estimated bandwidths.}\label{box-4}
\end{figure}

\begin{figure}[!htbp]
\centering
\includegraphics[width=7.62cm]{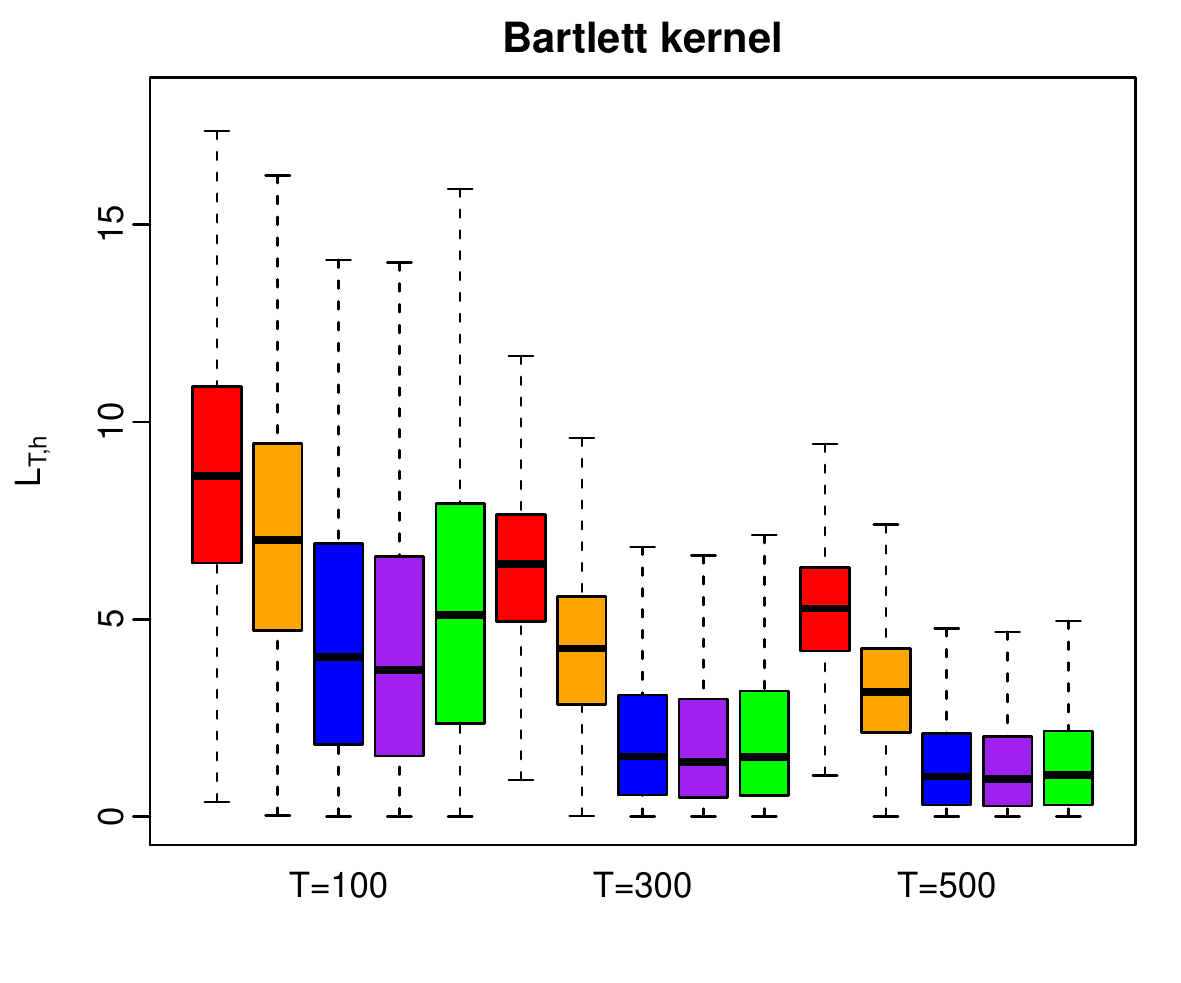}
\quad
\includegraphics[width=7.62cm]{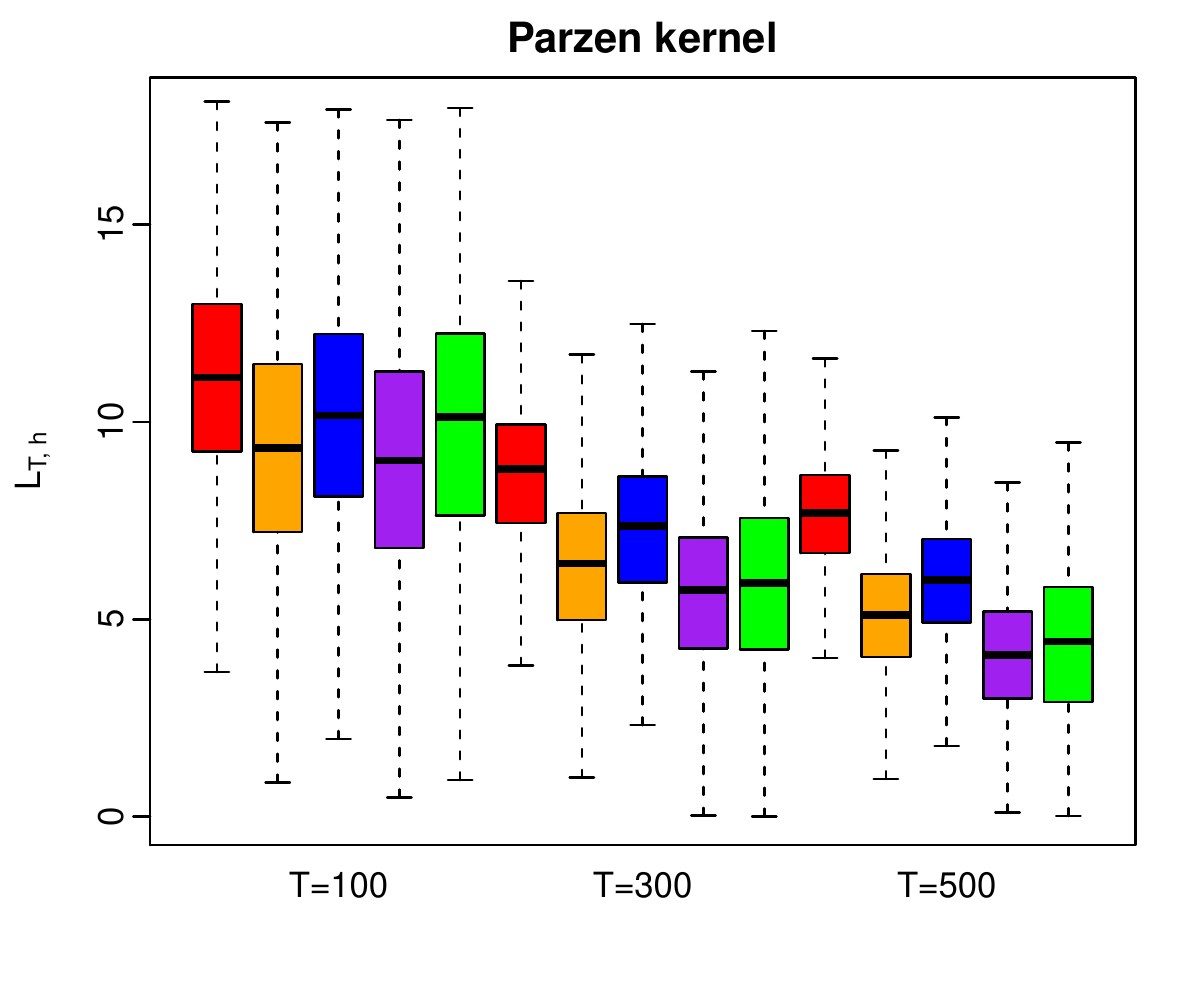}
\\
\includegraphics[width=7.62cm]{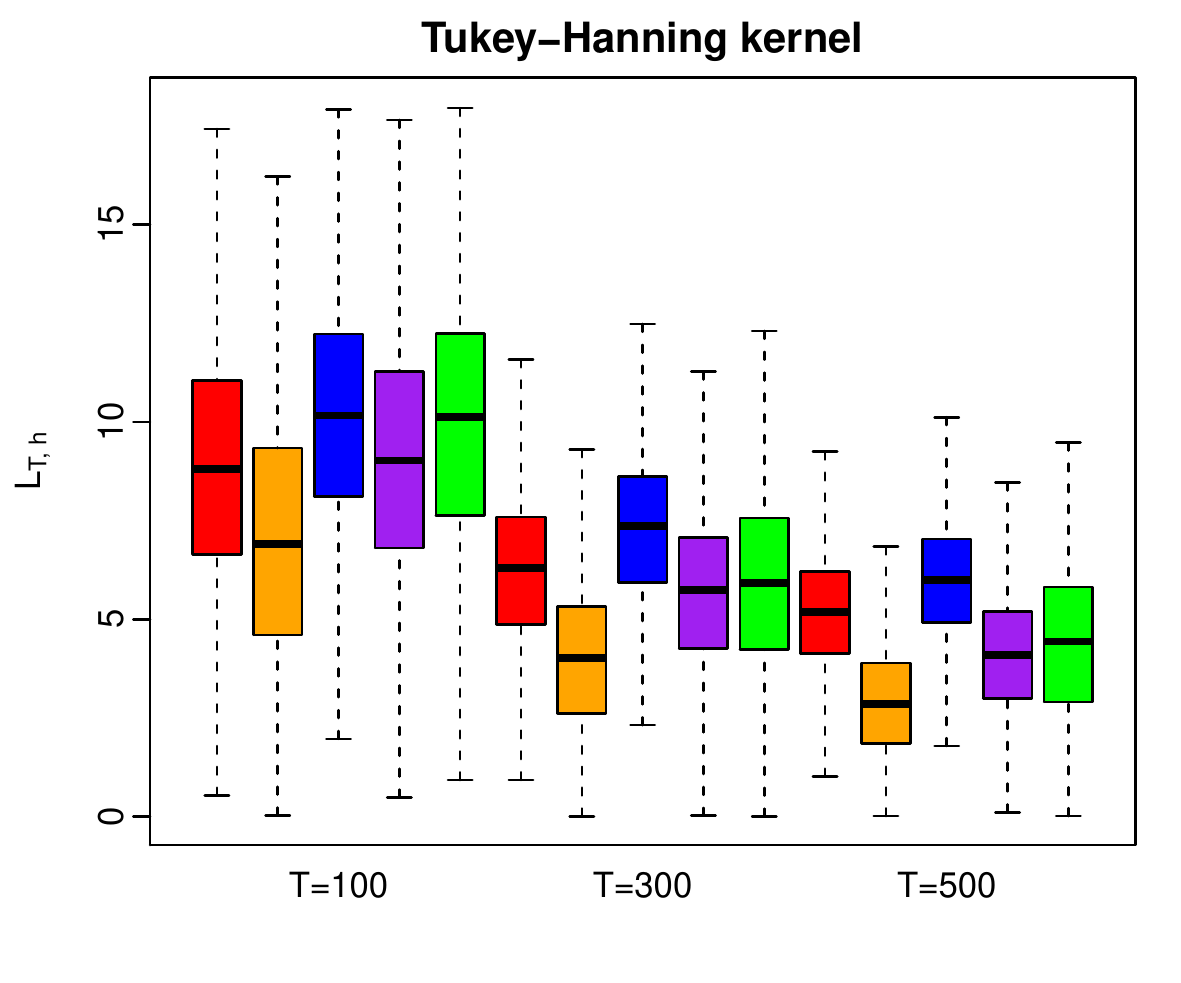}
\quad
\includegraphics[width=7.62cm]{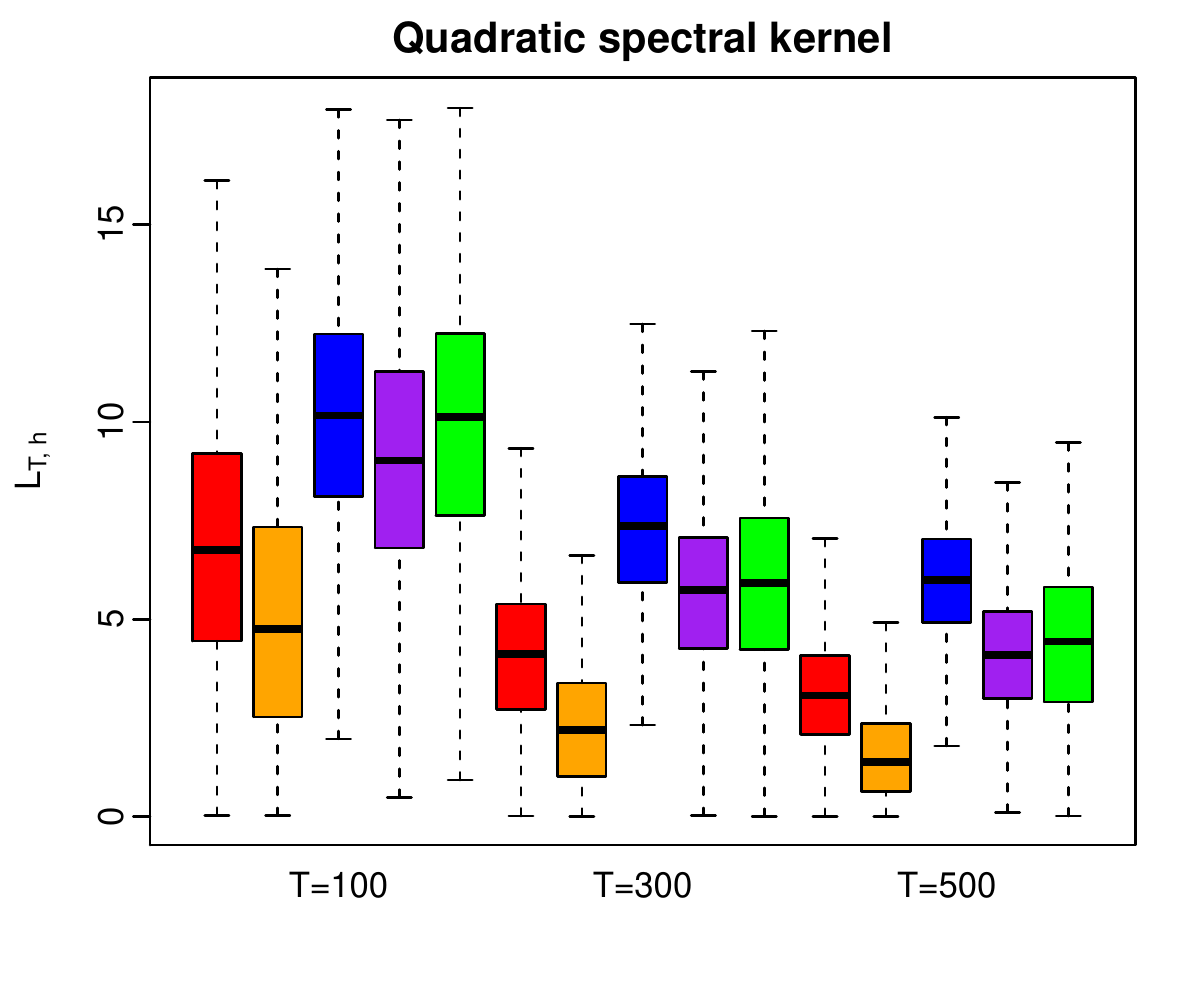}
\caption{Results for MA$_{\psi}(4)$ with estimated bandwidths.}\label{box-5}
\end{figure}

\begin{figure}[!htbp]
\centering
\includegraphics[width=7.62cm]{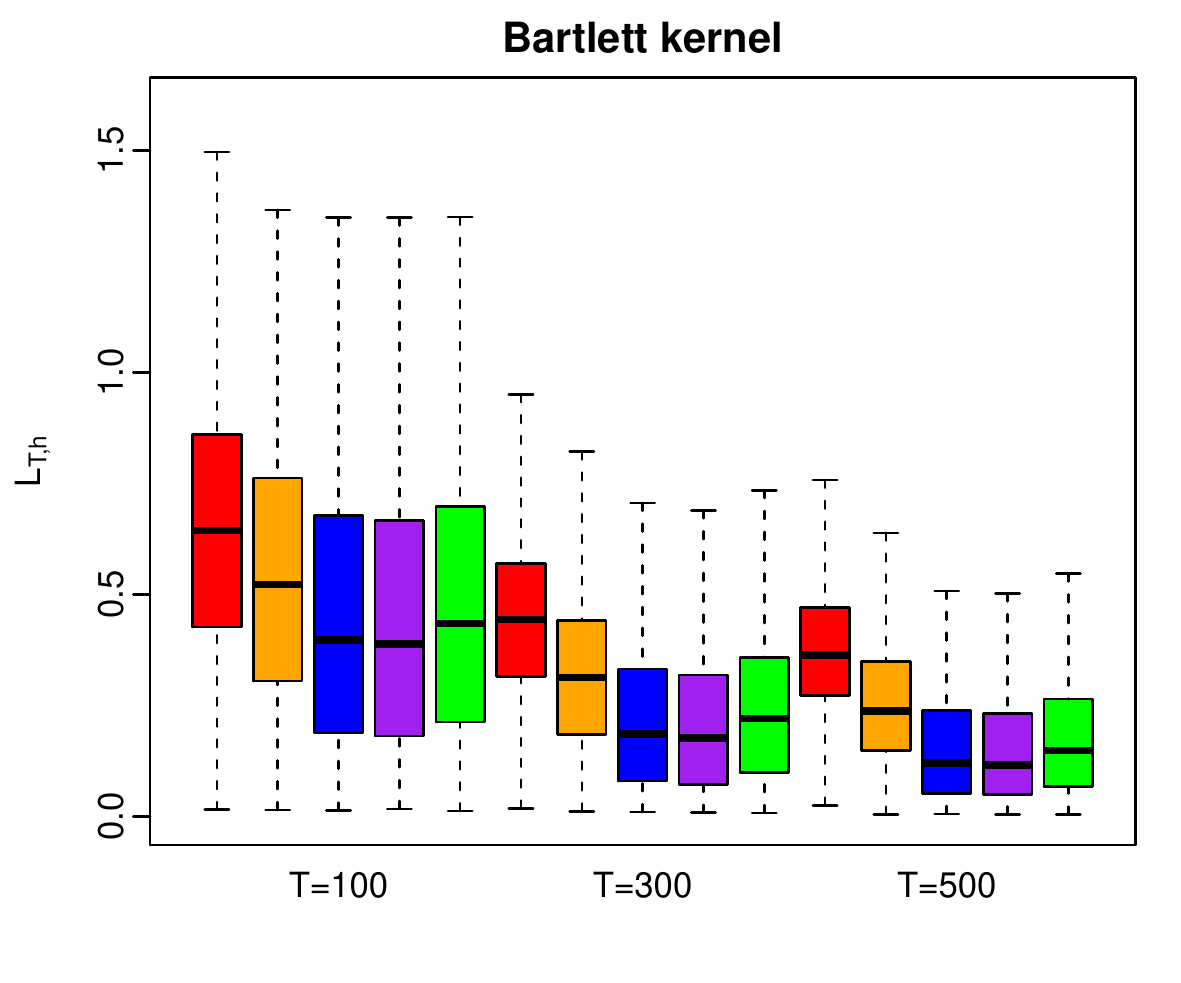}
\quad
\includegraphics[width=7.62cm]{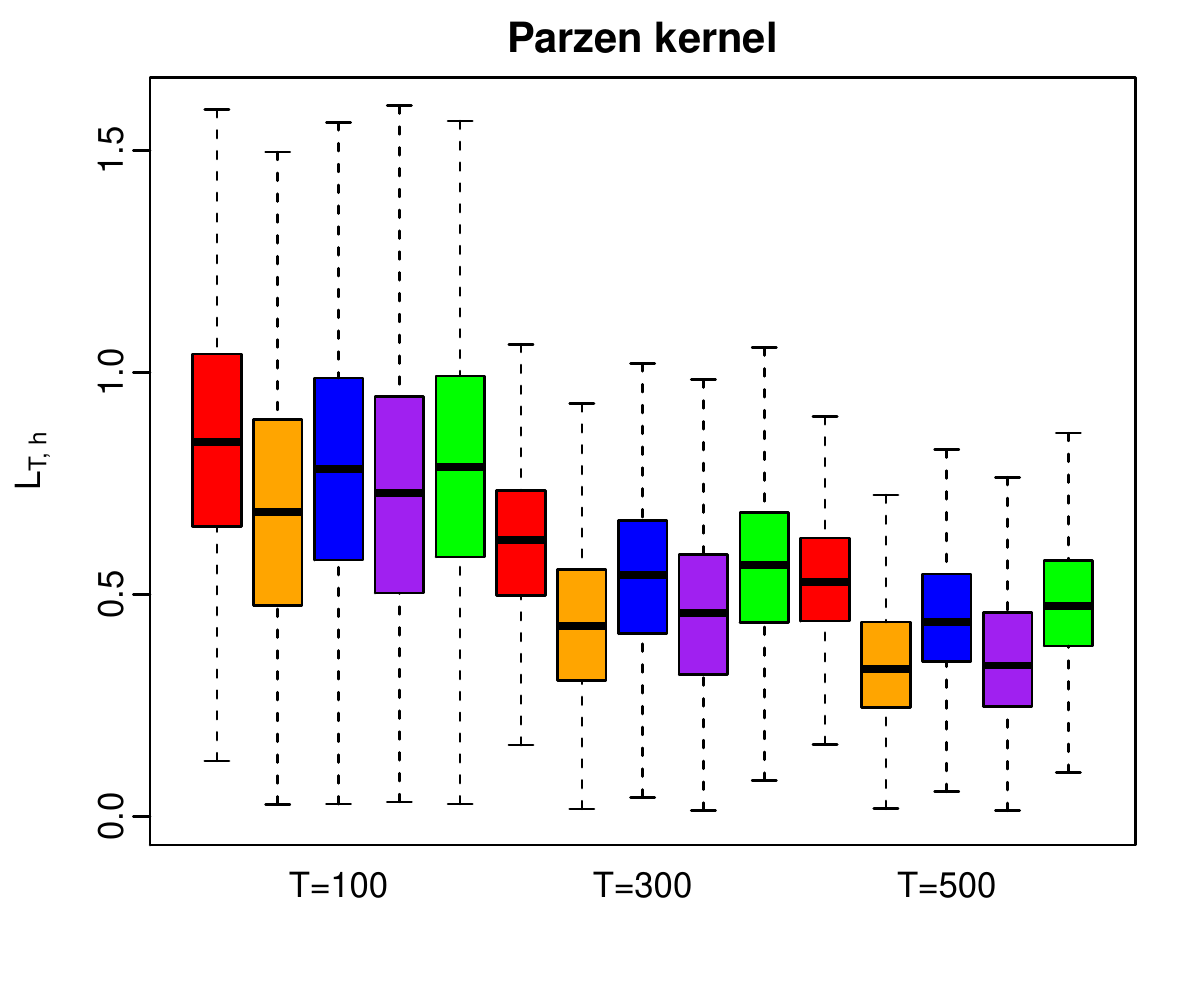}
\\
\includegraphics[width=7.62cm]{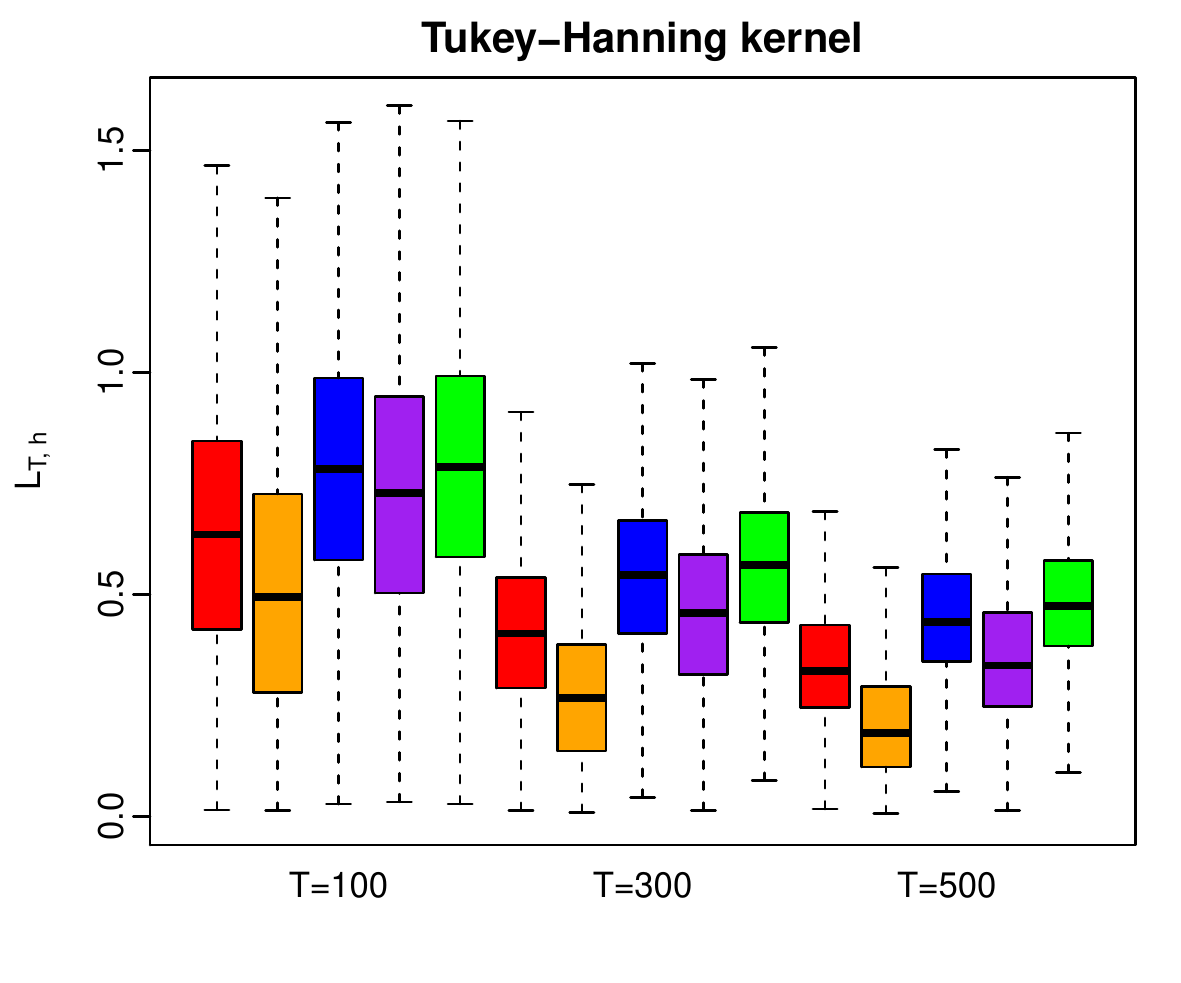}
\quad
\includegraphics[width=7.62cm]{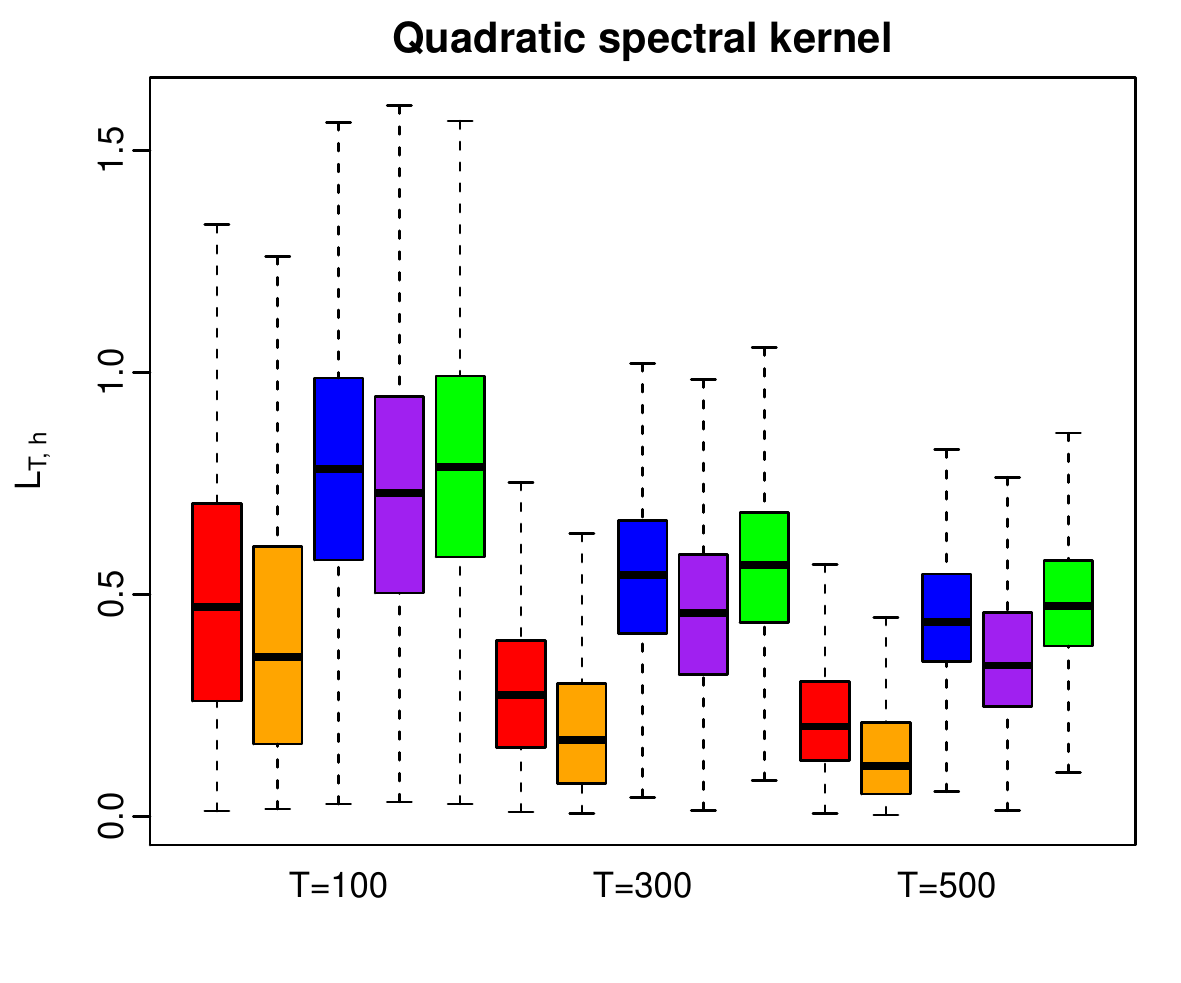}
\caption{Results for AR$_{0.5}(1)$ with estimated bandwidths.}\label{box-6}
\end{figure}

\begin{figure}[!htbp]
\centering
\includegraphics[width=7.62cm]{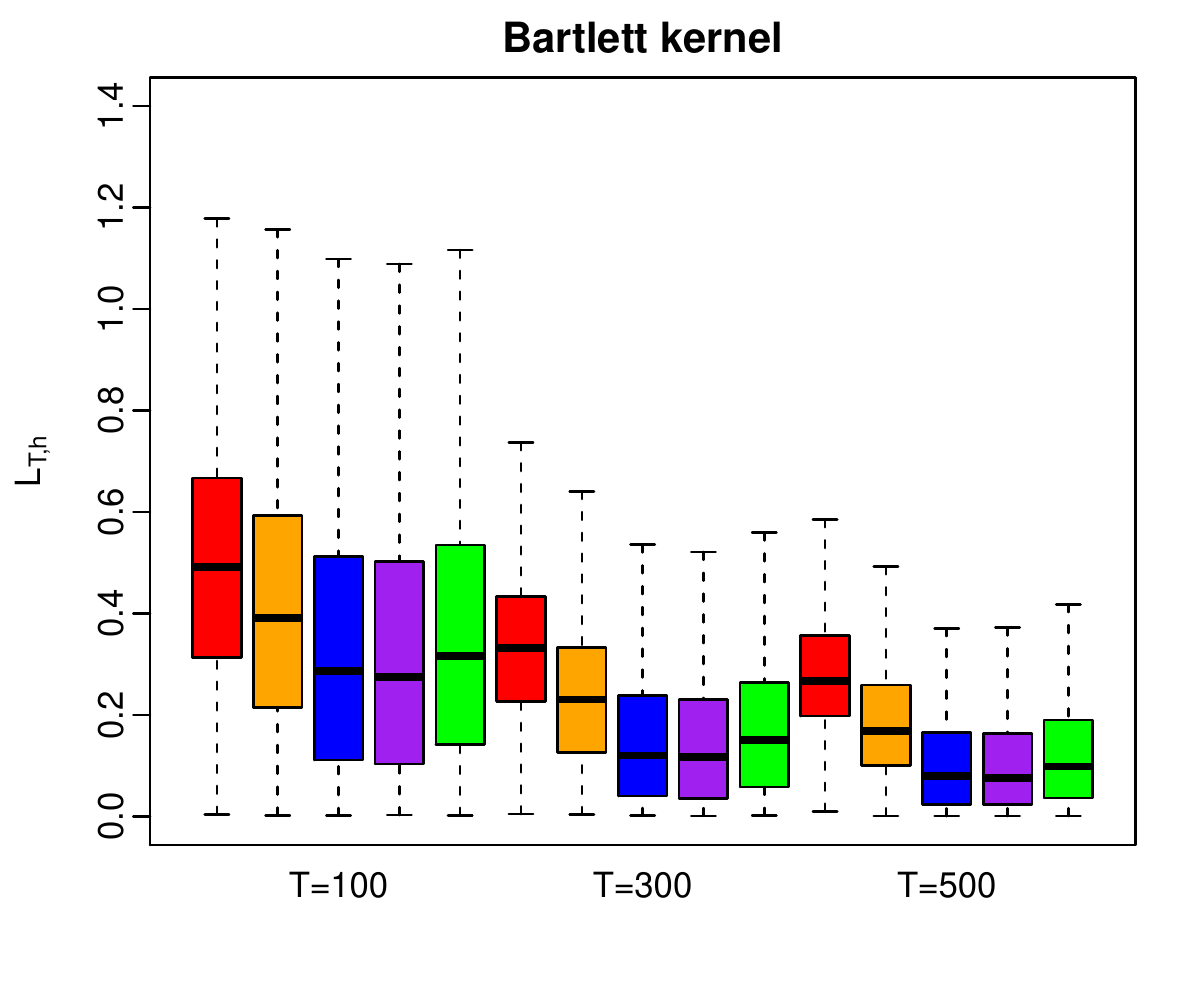}
\quad
\includegraphics[width=7.62cm]{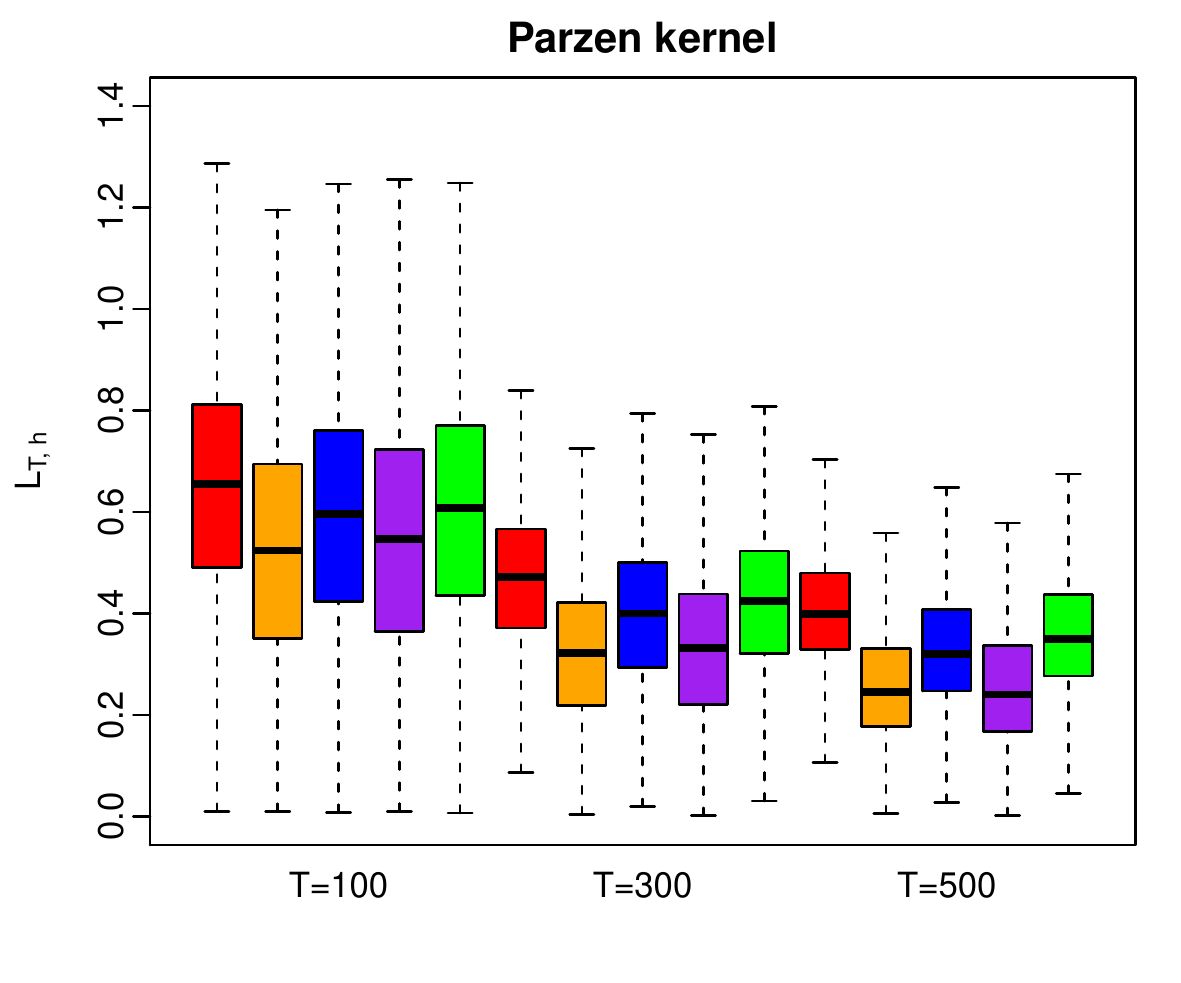}
\\
\includegraphics[width=7.62cm]{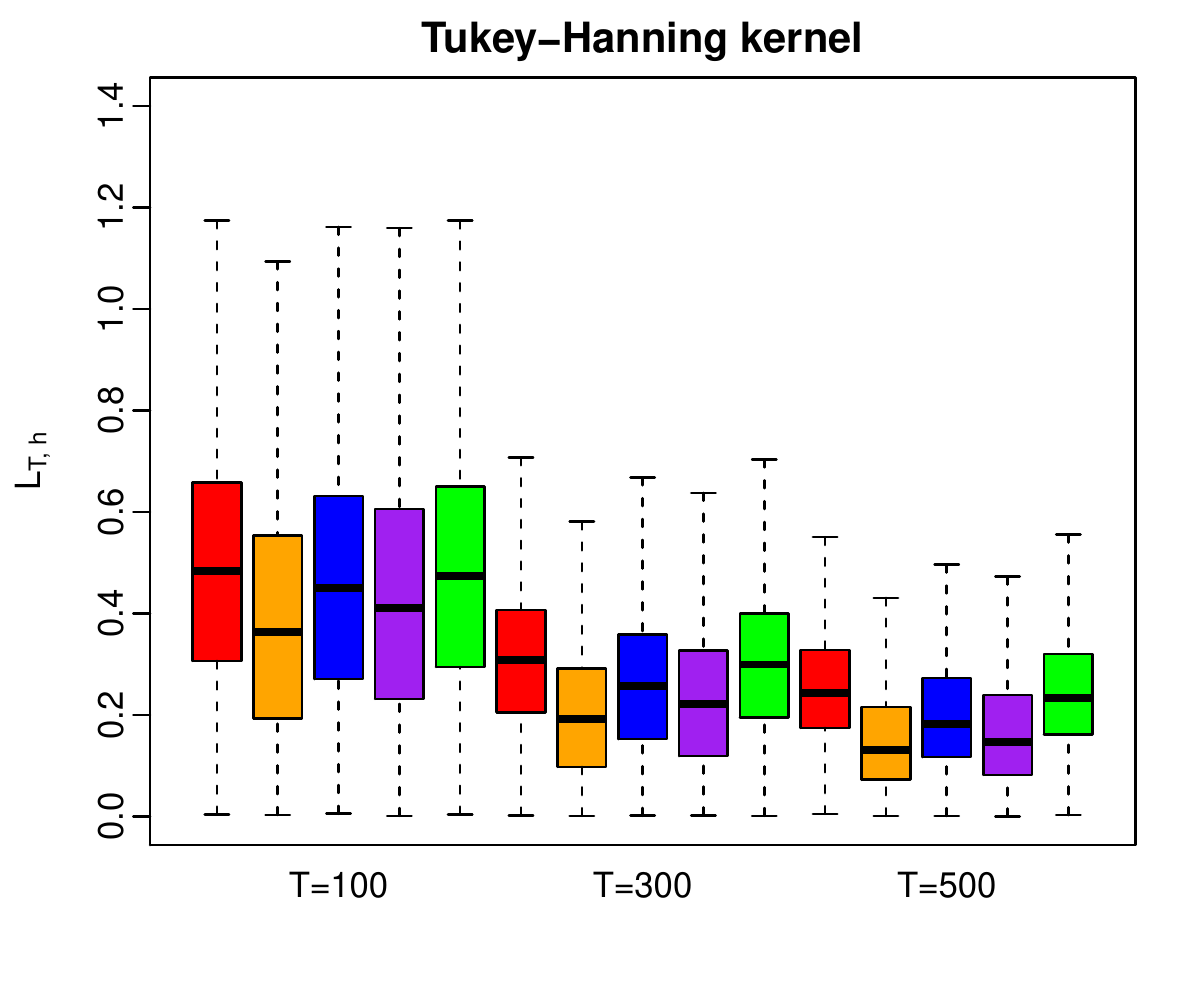}
\quad
\includegraphics[width=7.62cm]{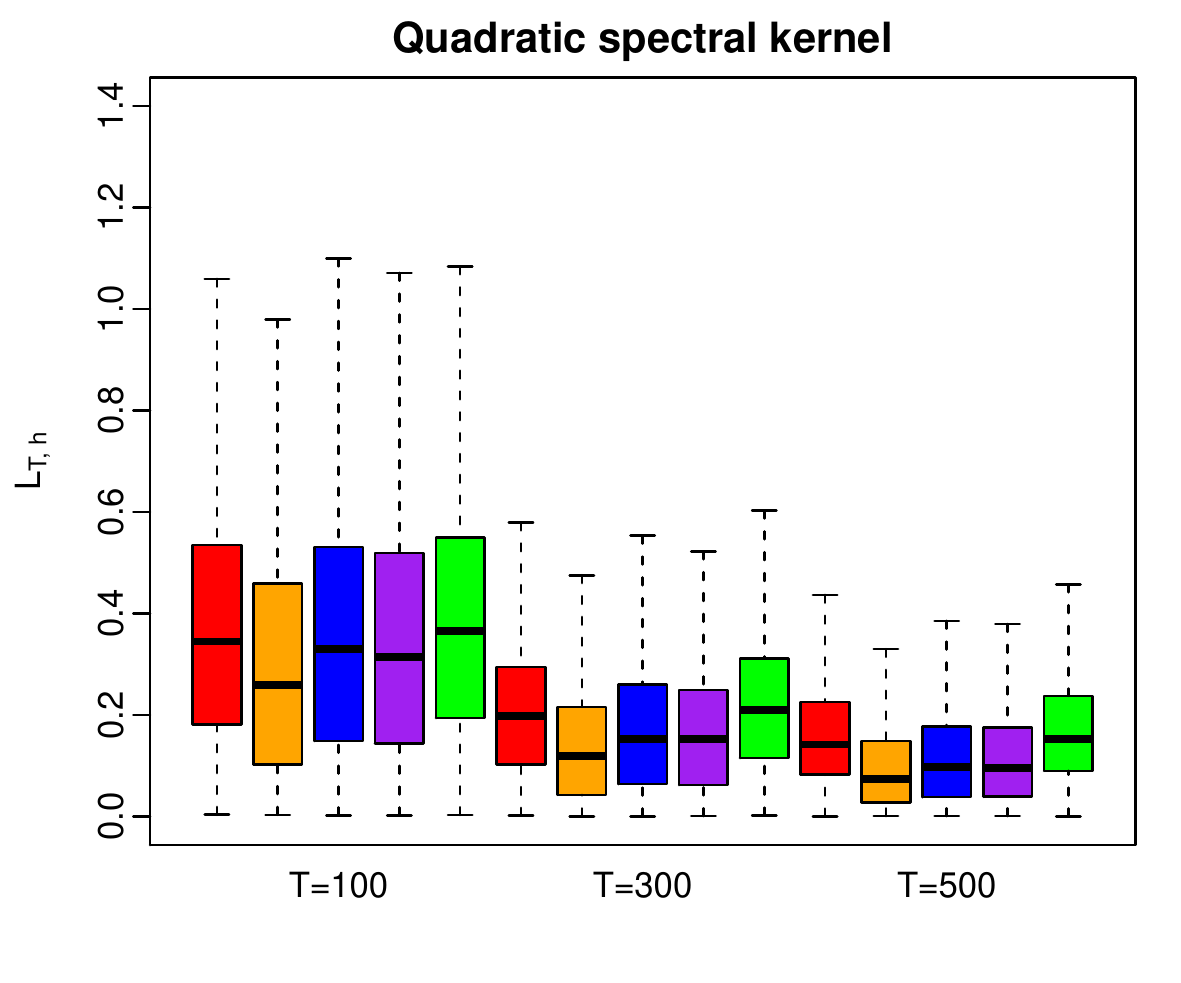}
\caption{Results for AR$_{\psi}(1)$ with estimated bandwidths.}\label{box-7}
\end{figure}

For each of 1000 repetitions of the Monte Carlo simulation over all weight functions, DGPs, and bandwidth selection settings outlined above, we approximate $L_{T,h} = \|\hat{C}_{h,q} - C\|^2$ by a simple Riemann sum approximation. The simulated values of $L_{T,h}$ are reported for all of the settings considered using box-plots in Figures \ref{box-1} to \ref{box-7}, which are color coded to indicate which setting they pertain to, and the legend for the color coding is given in the top left panel of Figure \ref{box-1}.  Based on these results, we draw the following conclusions:

\begin{itemize}
\item When the data are very weakly dependent (MA(0) and MA(1) cases) and the sample size is small, then the simple fixed bandwidths of $h=T^{1/5},T^{1/4}$ coupled with the Q-S weight function performed the best, although the improvements over the plug-in method are only modest in this case. When the sample size is very large (T=500), then the plug-in bandwidths exhibited competitive, and sometimes stronger, performance, although again only modestly better than the fixed bandwidths.
\item When the level of dependency is moderate to high (MA(4), MA(8), and FAR(1) cases), one of the plug-in bandwidth settings (Settings 3, 4, and 5) coupled with a Bartlett weight function exhibited the best performance in terms of the median values $L_{T,h}$ across all $T$. This interestingly coincides with Theorem \ref{th-1}, which suggests choosing the final order $q$ to be small in order to improve the optimal bandwidth estimation accuracy. The estimation accuracy improvements are again only modest when the dependence is moderate, especially compared to taking $h=T^{1/4}$ and the Q-S weight function, but fairly substantial accuracy gains are possible when the data is strongly dependent.

{\it Recommendation:}

\item When estimating $C$ as a part of an inferential procedure, our overall recommendation is to compare the results using two different estimates: 1) taking $h=T^{1/4}$ and the Q-S weight function, and 2) Using the plug in bandwidth selection procedure of Setting 4, which employs the flat-top weight function in the pilot estimation stage, coupled with a final Bartlett weight function estimate. One can expect only modest improvements in the overall estimation of $C$ when the functional time series data are close to being uncorrelated compared to this simple fixed bandwidth approach, however substantial accuracy gains can be achieved by the proposed method when the data are strongly dependent.
\end{itemize}

\section{Proofs} \label{proofs}

Throughout these proofs we let $c_i$, $i \ge 0$ denote unimportant numerical constants.

\begin{lemma}\label{lem-2} If Assumptions \ref{stat-ergo} and \ref{as-r}($r$) hold, and $h=h(T) \to \infty$,  then
\begin{align}\label{approx1}
\left| \| \hat{C}^{(p)}_{h,q_1} \| - \|C^{(p)}\| \right| = O_P\left( \left(\frac{h^{2p+1}}{T}\right)^{1/2} + h^{-\alpha} \right),
\end{align}
for all $0 \le p \le q$, and
\begin{align}\label{approx2}
\left| \int \hat{C}_{h,q_1}(u,u)du - \int C(u,u)du \right| = O_P\left( \left(\frac{h}{T}\right)^{1/2} + h^{-\alpha} \right),
\end{align}
where $\alpha= \min\{q_1,r\}$.

\begin{proof}
We begin by establishing \eqref{approx1}. According to the triangle inequality,
\begin{align}\label{a-1-1}
\left| \| \hat{C}^{(p)}_{h,q} \| - \|C\| \right| \le \| \hat{C}^{(p)}_{h,q} - C \|.
\end{align}
Under Assumptions \ref{stat-ergo} and \ref{as-r}, we obtain from equation (2.15) in Berkes et al. (2016) that, in the case when $p=0$,

\begin{equation*}
E\| \hat{C}_{h,q} - C \|^2 = O\left(\frac{h}{T} + h^{-2\alpha}\right).
\end{equation*}
Hence Chebyshev's inequality implies that
\begin{equation*}
\| \hat{C}_{h,q} - C \| = O_P\left( \left(\frac{h}{T}\right)^{1/2} + h^{-\alpha} \right),
\end{equation*}
which along with \eqref{a-1-1} implies \eqref{approx1} in this case. The approximation in \eqref{approx2} follows in an analogous manner as above which we outline below. According to Jensen's and Lyapounov's inequalities,

\begin{align*}
\left| \int \hat{C}_{h,q}(t,t)dt - \int C(t,t)dt \right| &\le \int |\hat{C}_{h,q}(t,t) -  C(t,t)|dt \\
&\le \left( \int |\hat{C}_{h,q}(t,t) - C(t,t)|^2 dt\right)^{1/2}.
\end{align*}
Moreover,
\begin{align*}
E\int |\hat{C}_{h,q}(t,t)- C(t,t)|^2dt &= \int {\rm var}(\hat{C}_{h,q}(t,t))dt + \int (E\hat{C}_{h,q}(t,t)-C(t,t))^2dt \\
& = O\Biggl(\frac{h}{T} + h^{-2\alpha}\Biggl),
\end{align*}
by Lemma 4.3 and Theorem 2.3 of Berkes et al (2016), which implies \eqref{approx2}. Towards establishing the rest of \eqref{approx1} for $0< p \le r$, one has that the integrated mean squared error of $\hat{C}^{(p)}_{h,q}$ may be broken into a variance and bias term as follows:
\begin{align}\label{f-mse}
E\| \hat{C}^{(p)}_{h,q} -{C^{(p)}}\|^2 = \intt {\rm var}( \hat{C}^{(p)}_{h,q}(u,s))duds + \|E\hat{C}^{(p)}_{h,q} - {C^{(p)}}\|^2.
\end{align}
According to the definition of $\hat{C}^{(p)}_{h,q}(u,s)$,
\begin{align*}
{\rm var}( \hat{C}^{(p)}_{h,q}(u,s)) &= \sum_{g,\ell=-\lfloor c h \rfloor }^{\lfloor c h \rfloor} W_{q_1}(g/h)W_{q_1}(\ell/h) |g|^p |\ell|^p {\rm cov}(\hat{\gamma}_g(u,s),\hat{\gamma}_\ell(u,s)) \\
&\le c^2h^{2p}\sum_{g,\ell=-\lfloor c h \rfloor }^{\lfloor c h \rfloor} W_{q_1}(g/h)W_{q_1}(\ell/h)  {\rm cov}(\hat{\gamma}_g(u,s),\hat{\gamma}_\ell(u,s)).
\end{align*}
By Lemma 4.3 in \cite{berkes:horvath:rice:2016},
\begin{align*}
\intt \sum_{g,\ell=-\lfloor c h \rfloor }^{\lfloor c h \rfloor} W_{q_1}(g/h)W_{q_1}(\ell/h)  {\rm cov}(\hat{\gamma}_g(u,s),\hat{\gamma}_\ell(u,s))duds = O\left(\frac{h}{T}\right),
\end{align*}
and therefore
\begin{align}\label{var-f}
\intt {\rm var}( \hat{C}^{(p)}_{h,q}(u,s))duds = O\left(\frac{h^{2p+1}}{T}\right).
\end{align}
Furthermore, since $E\hat{\gamma}_\ell(u,s) = (1-|\ell|/n)\gamma_\ell(u,s)$,
\begin{align}\label{f-bias-d}
E&\hat{C}^{(p)}_{h,q}(u,s) - {C^{(p)}}(u,s) = \sum_{\ell=-\lfloor c h \rfloor }^{\lfloor c h \rfloor} W_{q_1}\left(\frac{\ell}{h}\right) |\ell|^p \left(1-\frac{|\ell|}{T}\right)\gamma_\ell(u,s)-  \sum_{\ell=-\infty }^{\infty}  |\ell|^p \gamma_\ell(u,s) \\
&= \sum_{\ell=-\lfloor c h \rfloor}^{\lfloor c h \rfloor} \left(W_{q_1}\left( \frac{\ell}{h}\right)-1\right)|\ell|^p \gamma_\ell(u,s) - \sum_{\ell >\lfloor c h \rfloor} |\ell|^p \gamma_\ell(u,s) - \sum_{\ell=-\lfloor c h \rfloor}^{\lfloor c h \rfloor} W_{q_1}\left(\frac{\ell}{h}\right)\frac{|\ell|^{p+1}}{T} \gamma_\ell(u,s) \notag \\
&=G_1(u,s) - G_2(u,s) - G_3(u,s).  \notag
\end{align}
By utilizing the triangle inequality, we get that
\begin{align}\label{g-1}
\|G_1\| &\le \sum_{\ell=-\lfloor c h \rfloor}^{\lfloor c h \rfloor} \left(W_{q_1}\left( \frac{\ell}{h}\right)-1\right)|\ell|^p \|\gamma_\ell\| \\
&=  h^{-\alpha} \sum_{\ell=-\lfloor c h \rfloor}^{\lfloor c h \rfloor} \left(\frac{|\ell|}{h}\right)^{-\alpha}\left(W_{q_1}\left( \frac{\ell}{h}\right)-1\right)|\ell|^{p+\alpha} \|\gamma_\ell\|. \notag
\end{align}
According to \eqref{order}, $\sup_{-\infty < x < \infty}x^{-s}(W_{q_1}(x)-1)=O(1)$ for all $0\le s \le q_1$. It then follows from Assumption \ref{as-r} that
\begin{equation*}
\sum_{\ell=-\lfloor c h \rfloor}^{\lfloor c h \rfloor} \left(\frac{|\ell|}{h}\right)^{-\alpha}\left(W_{q_1}\left( \frac{\ell}{h}\right)-1\right)|\ell|^{p+\alpha} \|\gamma_\ell\|=O(1),
\end{equation*}
and thus by \eqref{g-1},
\begin{align}\label{g-1-bound}
\|G_1\| = O(h^{-\alpha}).
\end{align}
By the triangle inequality and Assumption \ref{as-r},
\begin{align}\label{g-2-bound}
\|G_2\| &\le \sum_{\ell >\lfloor c h \rfloor} |\ell|^p \|\gamma_\ell\| \\
&\le c^{-r}h^{-r} \sum_{|\ell|\ge0} |\ell|^{r+p} \|\gamma_\ell\|=O(h^{-r}). \notag
\end{align}
Once again by the triangle inequality and Assumptions \ref{as-r} and \eqref{order},
\begin{align}\label{g-3-bound}
\|G_3\| &\le \sum_{\ell=-\lfloor c h \rfloor}^{\lfloor c h \rfloor} \frac{|\ell|^{p+1}}{T} \|\gamma_\ell\| \\
&\le \frac{h^{1-r}}{T} \sum_{\ell=-\infty }^{\infty} |\ell|^{p+r} \|\gamma_\ell\| = O\left(\frac{h^{1-r}}{T}\right). \notag
\end{align}

Combining \eqref{g-1-bound}-\eqref{g-2-bound} with \eqref{f-bias-d} and the triangle inequality gives
\begin{align}\label{f-bias-f}
\|E\hat{C}^{(p)}_{h,q} - {C^{(p)}}\|=O\left(h^{-\alpha} + \frac{h^{1-r}}{T}\right),
\end{align}
where $\alpha=\min\{r,q\}$. This along with \eqref{var-f} and \eqref{f-mse} give \eqref{approx1} for $p\ge 0$.
\end{proof}

\end{lemma}

\begin{lemma}\label{lem-1}
Let $\hat{h}_{opt}$ be defined in \eqref{hopt-def}. Then under the conditions of Theorem \ref{th-1},
\begin{align} \label{h-est}
\hat{h}_{opt}(h_1,q_1) = h_{opt}\left(1+ O_P(f(T,h_1,q_1))\right),
\end{align}
where
\begin{equation*}
f(T,h_1,q_1) = \max \left\{ \biggl| \|\hat{C}^{(q)}_{h_1,q_1}\| - \|C^{(q)}\| \biggl|, \biggl| \|\hat{C}_{h_1,q_1}\| - \|C\| \biggl|, \left| \int \hat{C}_{h_1,q_1}(u,u)du - \int C(u,u)du \right| \right\}.
\end{equation*}
and $\hat{C}^{(p)}_{h_1,q_1}$ appearing in the definition of $\hat{h}_{opt}$ are defined in \eqref{cp-def}.

\begin{proof}
Simple algebra yields that
\begin{equation*}
\hat{h}_{opt}(h) = h_{opt}\left(1+ \frac{\hat{h}_{opt}(h)-h_{opt}}{h_{opt}}\right),
\end{equation*}
and, according to the definitions of $h_{opt}$ and $\hat{h}_{opt}(h_1,q_1)$,

\begin{align}\label{h-eq-1}
\frac{\hat{h}_{opt}(h_1,q_1)-h_{opt}}{h_{opt}}= c_1\Biggl( \Biggl| \frac{  \|\hat{C}^{(q)}_{h_1,q_1}\| }{\|\hat{C}_{h_1,q_1}\|^2 + \big(\int \hat{C}_{h_1,q_1}(u,u)du\big)^2}& \Biggl|^{1/(2q+1)} \\
&- \Biggl| \frac{  \|C^{(q)}\| }{\|C\|^2 + \big(\int C(u,u)du\big)^2}\Biggl|^{1/(2q+1)} \Biggl).\notag
\end{align}

An application of the mean value theorem applied to the right hand side of \eqref{h-eq-1} yields that
\begin{align}\label{eq-l1}
\frac{\hat{h}_{opt}(h_1,q_1)-h_{opt}}{h_{opt}} \le c_1 \theta_{m}  \Biggl| \frac{  \|\hat{C}^{(q)}_{h_1,q_1}\| }{\|\hat{C}_{h_1,q_1}\|^2 + \big(\int \hat{C}_{h_1,q_1}(u,u)du\big)^2} - \frac{  \|C^{(q)}\| }{\|C\|^2 + \big(\int C(u,u)du\big)^2}\Biggl|,
\end{align}
where $\theta_{m}=O_P(1)$ due to Lemma \ref{lem-2} and the assumption that $h=o(T^{1/(2q+1)})$. Simple algebra along with the triangle inequality and another application of the mean value theorem shows that

\begin{align}
\Biggl| \frac{  \|\hat{C}^{(q)}_{h_1,q_1}\| }{\|\hat{C}_{h_1,q_1}\|^2 + \big(\int \hat{C}_{h_1,q_1}(u,u)du\big)^2} - \frac{  \|C^{(q)}\| }{\|C\|^2 + \big(\int C(u,u)du\big)^2}\Biggl| \le  f(T,h_1,q_1)(1+O_P(1)),
\end{align}
which with \eqref{eq-l1} implies the Lemma.
\end{proof}
\end{lemma}

{\it Proof of Theorem \ref{th-1}:} By Lemma \ref{lem-1}, we have that
\begin{equation*}
\hat{h}_{opt}(h_1,q_1) = h_{opt}\left(1+ O_P(f(T,h_1,q_1))\right).
\end{equation*}
Furthermore, it follows from Lemma \ref{lem-2} that
\begin{equation*}
f(T,h_1,q_1)=  O_P\left( \left(\frac{h_1^{2q+1}}{T}\right)^{1/2} + h_1^{-\alpha} \right),
\end{equation*}
so that when $h_1 = A T^{\kappa}$, with $A \ge 0$,
\begin{align*}
f(T,h_1,q_1)&=  O_P\Biggl( T^{\kappa ( q + 1/2)-1/2} + T^{-\alpha \kappa} \Biggl), \\
& =O_P( T^{-\beta} ),
\end{align*}
where $\beta= \min\{ 1/2-\kappa(q+1/2), \kappa\alpha \}$ is defined in the statement of the theorem.

\qed

\begin{lemma}\label{lem-3}
If Assumption \ref{as-e} holds, then for all $p\ge0$,
\begin{align*}
\|E\hat{C}^{(p)}_{h,\infty} - {C^{(p)}}\| = O(h^p e^{-k_1 d h}),
\end{align*}
and
\begin{align*}
\int |E\hat{C}_{h,\infty}(u,u) - C(u,u)| du  = O(h^p e^{-k_1 d h}),
\end{align*}
where $k_1$ is defined in \eqref{flattop}.
\begin{proof}
As in the derivation of \eqref{f-bias-d} above, we have that
\begin{align}\label{f-bias-d-2}
E\hat{C}^{(p)}_{h,\infty}(u,s) - {C^{(p)}}(u,s) =G_1(u,s) - G_2(u,s) - G_3(u,s),
\end{align}
where
\begin{align*}
G_1(u,s)&=\sum_{\ell=-\lfloor k_2 h \rfloor}^{\lfloor k_2 h \rfloor} \left(W_\infty \left( \frac{\ell}{h}\right)-1\right)|\ell|^p \gamma_\ell(u,s),\\
G_2(u,s)&=\sum_{\ell >\lfloor k_2 h \rfloor} |\ell|^p \gamma_\ell(u,s),
\end{align*}
and
\begin{equation*}
G_3(u,s)=\sum_{\ell=-\lfloor k_2 h \rfloor}^{\lfloor k_2 h \rfloor} W_\infty \left(\frac{\ell}{h}\right)\frac{|\ell|^{p+1}}{T} \gamma_\ell(u,s).
\end{equation*}
It follows similarly to \eqref{g-3-bound} that $\|G_3\|=O(1/T)$. Also, by the triangle inequality and Assumption \ref{as-e},
\begin{align}
\| G_2 \| \le \sum_{\ell >\lfloor k_2 h \rfloor} |\ell|^p \|\gamma_\ell\| &\le c_3 \sum_{\ell >\lfloor k_2 h \rfloor} |\ell|^p e^{-d|\ell|} \\
&= O(h^p e^{-d k_2 h}). \notag
\end{align}
With regards to $G_1$, we have using the definition of $W_\infty$ that
\begin{align}
\| G_1 \| &\le \sum_{\ell=-\lfloor k_2 h \rfloor}^{\lfloor k_2 h \rfloor} \Biggl|W_\infty \Biggl( \frac{\ell}{h}\Biggl)-1\Biggl||\ell|^p \|\gamma_\ell\| \\
 &=   \sum_{k_1 h \le |\ell| \le k_2h } \Biggl|W_\infty \Biggl( \frac{\ell}{h}\Biggl)-1\Biggl||\ell|^p \|\gamma_\ell\| \notag \\
 &\le  c_4 \sum_{k_1 h \le |\ell| \le k_2h } |\ell|^p e^{-d|\ell|} = O(h^p e^{-d k_1 h}), \notag
\end{align}
which completes the proof given \eqref{f-bias-d-2} and the triangle inequality. The remaining assertion of the lemma follows from similar calculation, and so we omit the details.

\end{proof}

\end{lemma}

{\it Proof of Theorem \ref{th-2}:} By Lemma \ref{lem-1}, we have that
\begin{equation*}
\hat{h}_{opt}(h_1,\infty) = h_{opt}\left(1+ O_P(f(T,h_1,\infty))\right).
\end{equation*}
We continue by establishing bounds for the three terms defining $f(T,h_1,\infty)$. We present the bound for the first term, and the same bounds for the remaining terms may be obtained similarly. As in \eqref{a-1-1},
\begin{align*}
\left| \| \hat{C}^{(q)}_{h_1,\infty} \| - \|C\| \right| \le \| \hat{C}^{(q)}_{h_1,\infty} - C \|,
\end{align*}
and
\begin{equation*}
E\| \hat{C}^{(q)}_{h_1,\infty} - C \|^2=  \intt {\rm var}( \hat{C}^{(q)}_{h_1,\infty}(u,s))duds + \|E\hat{C}^{(q)}_{h_1,\infty} - {C^{(q)}}\|^2.
\end{equation*}

By the same calculations used to establish \eqref{var-f}, we have that
\begin{align}\label{var-f}
\intt {\rm var}( \hat{C}^{(q)}_{h_1,\infty}(u,s))duds = O\left(\frac{h_1^{2q+1}}{T}\right).
\end{align}
Also, it follows from Lemma \ref{lem-3} that
\begin{align*}
\|E\hat{C}^{(q)}_{h_1,\infty} - {C^{(q)}}\| = O(h_1^q e^{-k_1 d h_1}).
\end{align*}

Therefore by Markov's inequality,
\begin{equation*}
\left| \| \hat{C}^{(q)}_{h_1,\infty} \| - \|C\| \right|  = O_P\left( \left( \frac{h_1^{2q+1}}{T}\right)^{1/2} + h_1^q e^{-k_1 d h_1} \right).
\end{equation*}
The same bounds can be obtained for the remaining two terms in the definition of $f(T,h_1,\infty)$, from which it follows that
\begin{equation*}
f(T,h_1,\infty)= O_P\left( \left( \frac{h_1^{2q+1}}{T}\right)^{1/2} + h_1^q e^{-k_1 d h_1} \right).
\end{equation*}
We get that for $h_1 = A log(T)$,
\begin{equation*}
f(T,h_1,\infty)= O_P\left(\frac{\log^{(2q+1)/2}(T)}{\sqrt{T}}\right),
\end{equation*}
which completes the proof.
\qed

\bibliographystyle{apalike}
\bibliography{onestepARXIV.bib
}

\end{document}